\documentclass[sn-basic]{sn-jnl}

\usepackage{graphicx}%
\usepackage{multirow}%
\usepackage{amsmath,amssymb,amsfonts}%
\usepackage{amsthm}%
\usepackage{mathrsfs}%
\usepackage[title]{appendix}%
\usepackage{xcolor}%
\usepackage{textcomp}%
\usepackage{manyfoot}%
\usepackage{booktabs}%
\usepackage{algorithm}%
\usepackage{algorithmicx}%
\usepackage{algpseudocode}%
\usepackage{listings}%

\theoremstyle{thmstyleone}%
\theoremstyle{thmstyletwo}%

\theoremstyle{thmstylethree}%

\newcommand{\mnras}{Mon. Not. R. Astron. Soc.} 

\begin{document}

\title[Dust Traps in Discs]{Models and Observational Predictions of Dust Traps in Protoplanetary Discs}


\author*[1]{\fnm{Paola} \sur{Pinilla}}\email{p.pinilla@ucl.ac.uk}

\affil*[1]{\orgdiv{Mullard Space Science Laboratory}, \orgname{University College London}, \orgaddress{\street{Holmbury St Mary}, \city{Dorking}, \postcode{RH5 6NT}, \state{Surrey}, \country{UK}}}


\abstract{This manuscript investigates the impact of key dust evolution parameters—disc viscosity, fragmentation velocity, and the initial gas disc critical radius—on dust retention and trapping in protoplanetary discs. Using models with and without pressure bumps, combined with radiative transfer simulations, images of the dust continuum emission at (sub-)millimeter wavelengths, their fluxes and observed disc sizes are presented. For discs without pressure bumps (smooth discs), significant dust mass can only be retained over Myr timescales when dust fragmentation velocities are low ($v_{\rm frag}=1$ m\,s$^{-1}$) and with viscosity  values of $\alpha=10^{-3}$. For such a combination of fragmentation velocity and viscosity the synthetic images show a bright inner emission follow by a shallow emission with potential gaps if they are present in the gas profile as well. At higher fragmentation velocities ($v_{\rm frag}=5$–10 m\,s$^{-1}$), most dust is lost due to radial drift at million-year timescales unless pressure traps are present, in which case dust masses can increase by orders of magnitude and structures are observed in synthetic images. The viscosity parameter strongly shapes observable features, with low $\alpha$ producing sharper, potentially asymmetric inner wall cavities in inclined discs due to optically thick emission. High $\alpha$ favors the appearance of shoulders around the predominant rings that dust trapping produces. However, distinguishing between different fragmentation velocities observationally remains challenging. The inferred dust disc sizes from synthetic observations do not always correspond directly to dust model sizes or to the location of pressure bumps. Finally, we discuss implications for pebble fluxes and the delivery of volatiles to the inner disc. These results emphasize the strong degeneracies among dust evolution parameters and highlight the need for multi-wavelength, high-resolution observations to disentangle the processes shaping the formation of planets and planetary embryos in protoplanetary discs.}




\maketitle

\section{Introduction}\label{sec1}

I was awarded the 2025 Price Medal of the Royal Astronomical Society in January 2025. A couple of months later, I received an invitation to write this article, which should be related to the topic of the award. This recognition was based on my work of dust traps in protoplanetary discs, which has been work that I have done standing on the shoulders of giants since I started my PhD in 2010. I have been fortunate to work with incredible collaborators around the world, who have contributed immensely to this topic in the last decades. In parallel, I have also been very privileged to live in a time when planet formation models can be tested in details by observations. This is thanks to the huge efforts of previous and current generations to develop powerful telescopes and instruments that allow us to study the regions where planets form with unprecedented detail. I am very humble to write this paper, in which I aim to show the importance of dust traps on planet formation and their observational diagnostics, which is a topic of constant growth in the last decades.

As a brief historical perspective, I  start by mentioning that our knowledge of the evolution of protoplanetary discs and planet formation started with the efforts for understanding the formation of our own Solar System. Early works on the formation of planets in our Solar System agreed that planets  form in a disc that results from angular momentum conservation of the gravitational collapse of a molecular cloud (or solar nebula) that forms a protostar \citep{alfven1954, lust1955, kuiper1956, hoyle1960, cameron1962}. Few years later, the first numerical  calculations of the dynamics and collapse of molecular clouds  for protostar formation were presented by \cite{larson1969, penston1969a, penston1969b} and \cite{shu1977}, and contemporaneously, disc viscous accretion theory was developed by \cite{lynden1974} and 
\cite{shakura1973}, the latter was focused on black hole accretion discs. These models are still vastly used in disc evolution theory, although the paradigm of what is the main driver of angular momentum transport in protoplanetary discs is one of the main open questions in the field, with magneto-hydrodynamical (MHD) winds \citep{blandford1982, ferreira1997, bai2013, tabone2022} and viscous evolution being two possible mechanisms.

Early in the development of the theory of planet formation, it was clear that if planets form in discs with similar conditions than the inter-stellar medium (ISM), the growth from (sub-)micron size dust particles to planets is one of the most complex problems in astrophysics, where the growth from micron-sized particles to kilometre-sized objects spans over 40 order of magnitude in mass. In this growth process, micron-sized particles that the disc inherits from the ISM during the collapse of the molecular cloud stick due to van der Waals forces \citep{dominik1997}. However, when the relative velocities are high, collisions can lead to the fragmentation of particles \citep[e.g.,][]{Benz2000, Blum2000}. This growth process, from micron-sized particles to planetesimals, and the dynamics of particles is controlled by the aerodynamical drag that dust particles feel within the disc. Hence, knowledge of the gas distribution is crucial to understand these first steps of planet formation. Due to the coupling of dust particles on to the gas, the velocity of the dust particles can originate from  Brownian motion, settling to the midplane, radial/azimuthal drift and turbulent motion \citep[][]{brauer2008, birnstiel2010, birnstiel2024}.

\begin{figure*}
    \centering
    \includegraphics[width=\textwidth]{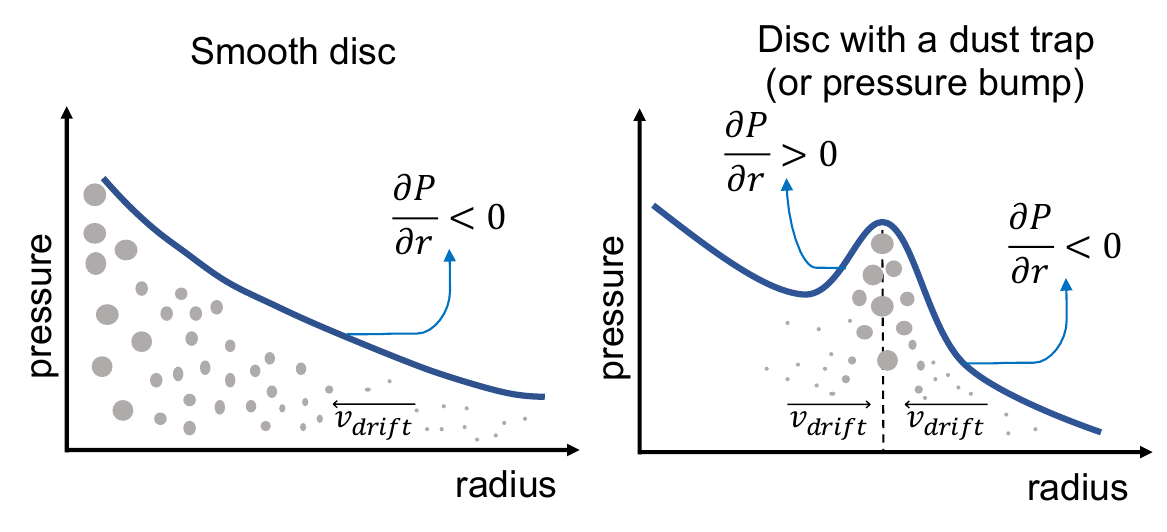}
    \caption{Sketch of the radial drift of particles in a smooth disc where $\partial_r P <0$ in the entire disc (left panel) vs. a disc where there is a pressure bump (right panel).}
    \label{fig:trap_cartoon}
\end{figure*}

One of the main barriers in the formation of pebbles  (millimeter to centimeter sized particles) and planetesimals  comes from the high radial drift velocities that particles experience when they are growing in the disc \citep{Wipple1972, Weidenschilling1977}. This radial drift originates from the head-wind that particles feel in the disc because the gas moves with slightly sub-Keplerian velocity when the disc pressure  decreases monotonically with radius (hereafter smooth disc). As a result, in  smooth discs, dust particles drift inwards (left panel of Fig.~\ref{fig:trap_cartoon}) and as explained below, these can lead to the rapid loss of pebbles towards the star before planetesimals and planets may form. 

In the Epstein drag regime, where  $a\lesssim (9/4)\lambda_{\mathrm{mfp}}$, with $\lambda_{\mathrm{mfp}}$ being  the mean free path of gas molecules, the drag of dust particles can be quantified by the  dimensionless Stokes number (St), which near and at the midplane is given by 

\begin{equation}
\textrm{St}=  \frac{a\rho_s}{\Sigma_g}\frac{\pi}{2},
\label{eq:stokes}
\end{equation}

\noindent where $a$ is the particle size, $\rho_s$ is the intrinsic volume density of the grains, usually assumed to be a few  g\,cm$^{-3}$, and  $\Sigma_g$ is the gas surface density. In response to the non-Keplerian gas motion, the dust particles move to regions of high pressure, with dust radial velocities given by \citep[e.g.,][]{Adachi1976, Nakagawa1986, Takeuchi2002, Youdin2010}

\begin{equation}
v_{\rm{r,\rm{dust}}}=\frac{v_{\rm{r,\rm{gas}}}}{1+\rm{St}^2}+ \frac{1}{\rm{St}^{-1}+\rm{St}} \frac{\partial_r P}{\rho\Omega}.
\label{eq:dust_vr}
\end{equation}

\noindent where $v_{\rm{r,\rm{gas}}}$ is the gas radial velocity,  $P$ is the disc pressure, $\rho$ is the disc gas volume density, and $\Omega$ the Keplerian frequency. 

The small particles with St$\ll1$, move with the gas accretion flow given by the first term in Eq.~\ref{eq:dust_vr}.  When the St number of particles increases, the second term in Eq.~\ref{eq:dust_vr} dominates and the radial velocity of particles is the highest  when St$=1$. In a smooth disc, where the pressure gradient is negative in the entire disc, the radial drift leads to the lost of particles towards the star. For typical disc parameters, particles with $\rm{St} = 1$ at the first astronomical unit correspond to $\sim50-100$\,cm sized-particles, which radial velocity would lead to lost towards the star in less than $\sim1000$\,yr \citep{brauer2008, pinilla_youdin2017}.

In potential regions where the pressure gradient could be positive, $ \partial_r P> 0$, dust particles experience outwards drift, and at pressure maximum, $\partial_r P=0$,  drift stops. As a consequence, pressure bumps are beneficial regions to reduce (around the pressure maximum) or completely halt (exactly at pressure maximum) the fast inward drift of particles and to overcome the radial drift barrier, as shown in the sketch of the right panel in Fig.~\ref{fig:trap_cartoon}. The accumulation of dust particles in pressure bumps, can lead to the formation of planetesimals, either by  gravitational collapse \citep{johansen2007} or by the streaming instability \citep{youdin2005}.

In a smooth disc, the radial drift of particles leads to the fast depletion of pebbles (St$\gtrsim 10^{-2}$), contradicting observations of protoplanetary discs. Specifically, multi-wavelength observations in (sub-)millimetre range suggest that grains in protoplanetary discs are larger than in the ISM, with millimetre /centimetre  sizes. This evidence comes from the slope of the spectral energy distribution (SED) at millimetre  wavelength or spectral index of protoplanetary discs ($\alpha_{\rm{mm}}$), which has values lower than the ISM value of 3.5-3.7, independent of the star forming region that they belong to (or age) or stellar type host \citep[e.g.,][]{testi2003, testi2014, wilner2005, rodmann2006, ricci2010, ricci2012, pinilla2017b, tazzari2021, kurtovic2025}. 

\cite{pinilla2012} suggested that the discrepancy between the models including radial drift and observations could be solved by including global pressure bumps in the disc. In these models, static pressure bumps are assumed to live in the disc for million-year timescales. Models with pressure bumps with an amplitude of $\sim$30\% with respect to the unperturbed background density (or pressure) profile can explain the observed millimeter spectral indices and the survival of pebbles in discs at different timescales of evolution ($<$1 - 10\, Myr). Later works of dust evolution assuming pressure bumps that are not static in the disc, and instead they slowly appear during the disc evolution and/or appear and disappear with time, demonstrated that pressure bumps need to be long-lived, and their onset needs to be at early times (before $\sim$1\,Myr) in order to explain observations, in particular the observed spectral indices \citep{stadler2022, delussu2024}. 

Fortunately, with the advance of recent telescopes, such as the Atacama Large Millimeter Array (ALMA), protoplanetary discs have been observed with unprecedented resolution (of few astronomical units) and sensitivity, revealing structures that can be the result of pressure bumps in the disc \citep[e.g.,][]{dullemond2018, rosotti2020}, and proving that they can be the solution of the long-standing barrier of radial drift. Apart from halting the radial drift, pressure bumps help to prevent that dust particles experience high relative velocities due to drift, potentially leading to more efficient growth.

\begin{figure*}
    \centering
    \includegraphics[width=\textwidth]{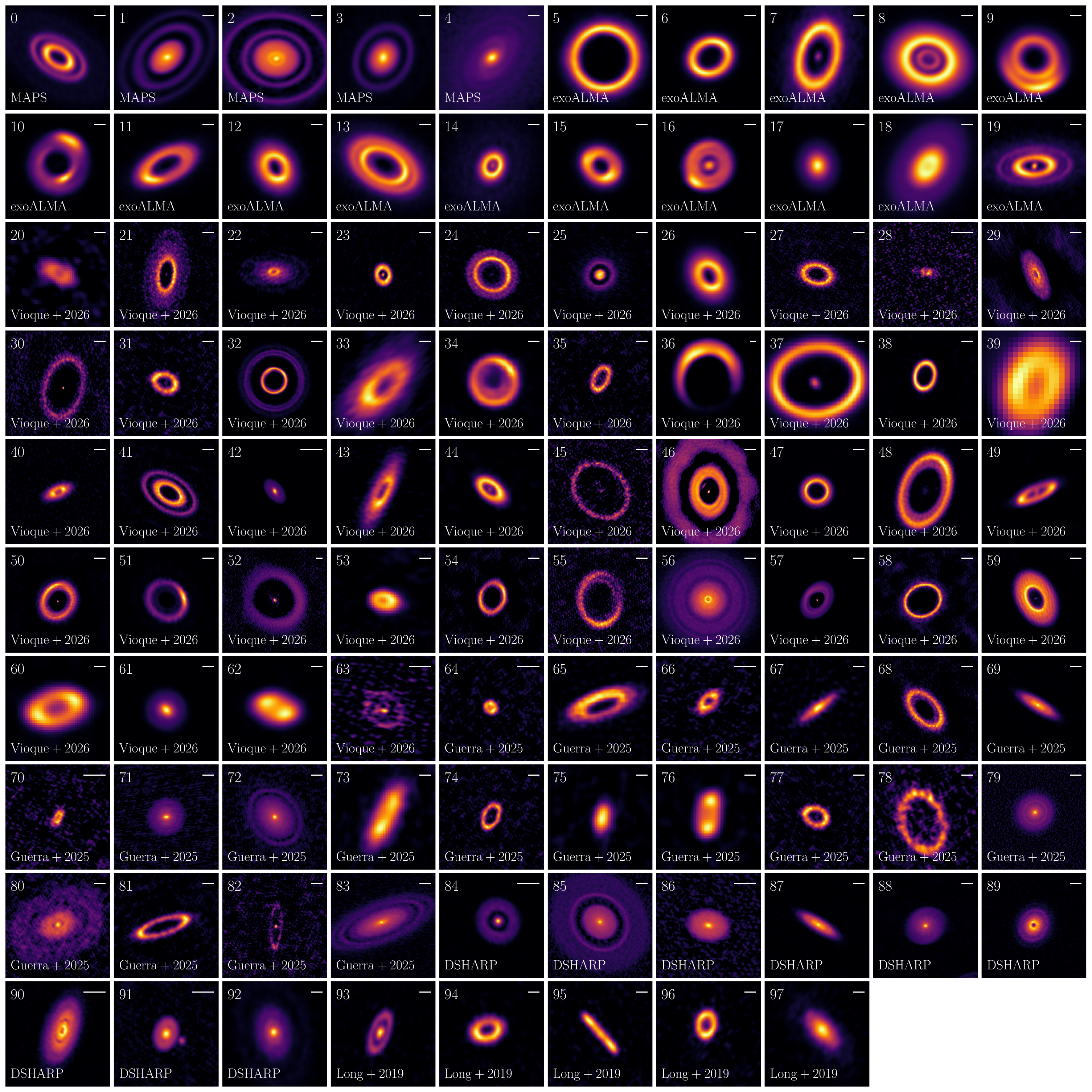}
    \caption{Examples of ALMA observations of protoplanetary discs with structures from different observing programs, including \citet[][DSHARP]{andrews2018}, \cite{long2019}, \citet[][MAPS]{Oberg2021}, \citet[][exoALMA]{Teague2025}, \cite{guerra2025}, and \citet[][and citations within]{Vioque2026}. In the upper right of each panel, a bar of 0.1'' is shown as a reference for the scale.}
    \label{fig:gallery}
\end{figure*}

Figure~\ref{fig:gallery} shows examples from ALMA observations of Class II protoplanetary discs (ages between $\sim$1-10\,Myr) from different observational programs. These observations are mostly at Band 6 and 7, and they reveal several structures, which could be the result of pressure bumps in discs and retaining the dust particles in specific regions, such as the observed rings. Structures have been also been detected in younger discs that are still embedded \citep[][]{SeguraCox2020, Ohashi2023, vioque2025}, but the detection of structures at such early phases remains challenging due to the high disk optical depth. ALMA observations, such as the collection in Fig.~\ref{fig:gallery}, have triggered a lot of work in the community in the last decade, to understand what is the origin of structures and what are the implications for planet formation \citep[e.g., ][]{dong2015, dong2018, kanagawa2015, Flock2015, pinilla2015, pinilla2017, zhang2018, garate2021, garate2023, bae2023}. 

In this  manuscript, I briefly present examples of dust evolution models with and without pressure bumps in the disc, and I explore how the efficiency of trapping depends on key parameters of dust evolution models, such as the dust fragmentation velocity and disc viscosity. In addition, I discuss the implications of dust traps on the pebble flux into the inner disc, which has recently gained attention due to the accumulated evidence that the molecular abundances in the first astronomical units revealed with the  James Webb Telescope (JWST) of several protoplanetary discs may be connected to the outer dust structure \citep[e.g.,][]{Arulanantham2025, gasman2025, krijt2025}. These inner disc chemical  abundances determine the type of material and the timescales that is available for the formation of (terrestrial) planets in the disc.

In addition, I present different observational diagnostics of pressure bumps in discs, in particular at the millimeter regime, including the potential detection of structures, their shape at different wavelengths, and how it affects dust disc sizes, fluxes, and spectral indices. I finish this work by discussing how observations can be used to better understand dust evolution and trapping, and to potentially distinguish the origin of pressure bumps in discs. 

\section{Models}

\subsection{Dust Evolution Models}

When I started my PhD, my colleague Til Birnstiel was finishing his PhD in the same research group, who had developed gas and dust evolution models that have been widely used by the community. In these models, the gas viscously evolves under the $\alpha-$ parameter prescription \citep{shakura1973}, while the dust growth, coagulation, and dynamics are included in the models \citep{birnstiel2010}. These models were based on the work by Frithjof Brauer \citep{brauer2008}, who passed away in 2009 and who I never had the honour to meet, but to whom I am very thankful for providing the baseline of the research that have been done in the topic of dust evolution and pressure bumps. 

Since I started my PhD, I have been working on the models developed in \cite{birnstiel2010} to include the influence of pressure bumps and to understand that are the observational diagnostics of them. Nowadays, there is a \texttt{python} package of these dust evolution models call  \texttt{Dustpy} introduced in \cite{stammler2022},  which improved several aspects of the \cite{birnstiel2010} models and it allows easy modifications or extensions to add different physics, such as the examples that are given below.

In short, \texttt{Dustpy} is a viscous gas evolution 1D (radial direction) code,  that solves the Smoluchowski equation \citep{Smoluchowski1916} to model the growth and fragmentation of particles, while simultaneously modeling the dust dynamics, including advection, and diffusion for multiple particle sizes in the radial direction. As it is modular, several physical ingredients  have been included in recent years, as for example  mass loss due to internal and external photoevaporation \citep{garate2021, garate2024},  dynamical back-reaction of the dust onto the gas \citep{garate2020}, removal of dust mimicking the formation of planetesimals \citep{stammler2023}, infall material replenishing the disc \citep{zhao2025}, and  soon non-ideal MHD effects based on the models of \cite{delage2022}.

In this study, I use \texttt{Dustpy} to implement pressure bumps in the disc and study the effect of different model parameters on the trapping efficiency.  A disc around a Solar mass star is assumed and initially the gas is an unperturbed profile as \citep{lynden1974}

\begin{equation}\label{eq:lbp}
    \Sigma_g(r,t_0) = \Sigma_0 \left(\frac{r}{R_\mathrm{c}}\right)^{-\gamma}\,\exp\left[-\left(\frac{r}{R_\mathrm{c}}\right)^{2-\gamma}\right],
\end{equation}

\noindent with $\gamma=1$. The value of $\Sigma_0$ is such that the gas disc mass is 0.05\,$M_\odot$. Two different characteristic radius $R_c$ are assumed in the models: 20 and 80\,au, representing discs that are compact or extended. In these cases, either all the assumed traps are inside  $R_c$ (in the case of $R_c=80$\,au) or only one of them (in the case of $R_c=20$\,au), as explained later. 

A radial grid from 1\,au to 300\,au with  $n_\mathrm{r}=300$ logarithmically spaced cells. The initial dust-to-gas mass ratio is 0.01, and the grains initially have sizes between 0.1-1\,$\mu$m following a grain size distribution as the ISM grains, such that $\propto a^{-3.5}$ \citep{mathis1977}. In addition, three values of the fragmentation velocity are considered: 1, 5, and 10\,m\,s$^{-1}$. These values are based on laboratory experiments and numerical simulations of dust collisions, which have shown a large variety of values for the threshold of the fragmentation velocities of particles, i.e., values at which particles are expected to fragment after collision  \citep[e.g.,][]{kempf1999, Blum2000, blum2008, krause2004, paszun2006, wada2009, wada2011, gundlach2015, musiolik2016, musiolik2019, Pillich2021}.

The midplane temperature profile is described by a power law of a flaring passively irradiated disc, such that 
\begin{equation} \label{eq:temp_profile}
    T(r) = T_\mathrm{eff}\,\left(\frac{0.5 \,\varphi_\mathrm{irr}\, R_*^2}{r^2} \right)^{1/4},
\end{equation}

\noindent where the stellar effective temperature $T_\mathrm{eff}=5772\,\mathrm{K}$, with a stellar radius $R_\star=R_\odot$ and an irradiation angle of $\varphi_\mathrm{irr}=0.05$. 

To include the effect of pressure bumps, I add a perturbation in  the viscous $\alpha$ profile, that creates a gap-like structure in the gas surface density profile \citep{dullemond2018, stadler2022, garate2024, kurtovic2025}.
The perturbation is a Gaussian profile, such that:
\begin{equation} \label{eq_alpha_bump}
    \alpha'(r) = \alpha \times \left[1 + A_\textrm{gap} \exp\left(-\frac{\left(r - r_\textrm{gap}\right)^2}{2w_\textrm{gap}^2}   \right)\right],
\end{equation}

\noindent where $\alpha$ is the viscosity base value, and $A_\textrm{gap}$, $r_\textrm{gap}$, and $w_\textrm{gap}$ are the gap amplitude, location, and width, respectively. Two values of the viscosity $\alpha$-parameter in  Eq.~\ref{eq_alpha_bump} are assumed: $10^{-4}$ and $10^{-3}$. This parameter has a strong influence on the dust evolution models because it controls the dust turbulent velocities, the dust settling, and the dust diffusion. It is important to note that these simulations assume isotropic turbulence, implying that the $\alpha$ parameter that controls the turbulent velocities of particles, their radial diffusion and vertical stirring is considered to be the same value (either $10^{-4}$ or $10^{-3}$), and only the $\alpha$ viscous parameter that controls the gas evolution changes according to Eq.~\ref{eq_alpha_bump}. For the simulations of this work, the dust diffusion is taken as

\begin{equation} \label{eq_dust_diff}
   D_d=\frac{\alpha c_s h_g}{1+\rm{St}^2}.
\end{equation}

Protoplanetary discs can have non-isotropic turbulence, implying that turbulent velocities and the radial and vertical mixing of dust particles can be disconnected from what regulates the gas viscous evolution, and have different values in the radial vs. the vertical direction, as explored in \cite{pinilla2021}.

In the models with pressure bumps, I include three gaps with an amplitude $A=4$ and a width ($w_\textrm{gap}$) that corresponds to the local disc scale height. These three gaps are assumed to be located at 10, 40 and 70\,au, which implies that $w_\textrm{gap}$=0.5, 3, 6\,au, respectively. An amplitude of $A=4$ resembles planet masses of $\sim0.1-1\,M_{\rm{Jup}}$ depending on different disc parameters \citep{zhang2018, pinilla2020}. The value of $A$ controls the pressure gradient, and hence it has a direct effect on the amount of dust that can be trapped. Table~\ref{Table_Parameters} summarises the parameters used for the gas and dust evolution models presented in this study. The simulations span 0 to 5\,Myr of evolution.

\begin{table}
\caption{Assumed parameters for the gas and dust evolution models (upper part of the table) and radiative transfer models (lower part of the table)}
\label{Table_Parameters}
\begin{tabular}{l l l }
 \hline \hline
Quantity/Unit & Description & Value  \\
\hline
$M_\star$[$M_\odot$]& Stellar mass & 1.0  \\
$T_\star$[K] & Stellar effective temperature& 5772\\
$r_{\rm{in}}$ [au] & Inner radial boundary& 1.0 \\
$r_{\rm{out}}$ [au] & Outer radial boundary& 300 \\
$R_c$ [au]&Cut-off radius& 20, 80\\
$n_r$&Number of radial cells&300\\
$M_{\rm{disc}}$ [$M_\star$]&Disc mass&0.05\\
$\alpha$&Disc viscosity&$10^{-4}$, $10^{-3}$\\
$v_{\rm{frag}}$ [m\,s$^{-1}$]&Fragmentation velocity&1, 5, 10\\
$\rho_s$[g\,cm$^{-3}$] & Grain material density & 0.9\\
$A$ & Amplitude of pressure bumps & 0, 4\\
$r_{\rm{gap}}$ [au] & Location of gaps & 10, 40, 70\\ 
\hline
$n_\theta$ & Number of polar cells & 100\\
$n_\phi$ & Number of azimuthal cells & 64\\
$n_\phi$ & Number of cells for the wavelengths & 150 \\
$n_{\rm{photons}}$ & Number of photons & $10^{7}$ \\
$n_{\rm{photons, scatt}}$ & Number of photons & $5\times10^{6}$ \\
$\lambda_{\rm{min}}$ [$\mu$m] & Minimum wavelength& 0.1 \\
$\lambda_{\rm{max}}$ [$\mu$m] & Maximum wavelength& $10^{4}$ \\
$d$ [pc]& Distance to the source & $140$\\
\hline
\end{tabular}
\vspace{-3.0mm}
\end{table}

\subsection{Radiative Transfer Models}

In this study, the results from dust evolution models are combined to radiative transfer models to give observational diagnostics of dust traps at the millimeter regime. For this,  the code \texttt{RADMC3D} \citep{dullemond2012} is used. The opacity of each grain size is calculated using \texttt{optool} \citep{dominik2021}, and it is assumed porous spherical grains for the opacities, with a composition as in \cite{ricci2012}, this means: a vacuum volume fraction of 40\%, 10\% silicate, 20\% carbon, and 30\% water ice.

For these radiative transfer models, it is assumed that the central star is the source of irradiation, and it is a Solar-type star with the same effective temperature than assumed in the dust evolution models (Table~\ref{Table_Parameters}).

The volume dust density of each grain size that it is used as an input in the \texttt{RADMC3D} models is obtained from the dust evolution models, such that:

\begin{equation}
        \rho_d(R,z, \mathrm{St}) = \frac{\Sigma_d(R, \mathrm{St})}{\sqrt{2\,\pi}\,h_d (R, \mathrm{St})}\,\exp \left( -\frac{z^2}{2\,h_{\mathrm{d}}^2(R, \mathrm{St})} \right)\,,
        \label{eq:volume_density}
\end{equation}

\noindent where  $z = r\,\cos(\theta)$ and $R=r\,\sin(\theta)$, with $\theta$ being a polar angle. For the grid of the polar angle, 100 cells are assumed from 0 to $\pi$. In the azimuthal angle, 64 cells are assumed from 0 to $2\pi$, but this is useful only to produce the images, as the models are axisymmetric.

The dust scale height $h_{\mathrm{d}}$ is obtained for each particle with a given St number by \citep{youdin2007, birnstiel2024}

\begin{equation}
        h_d(\mathrm{St})=h_g \sqrt{\frac{\alpha}{\rm{St}+\alpha}}
        \label{eq:dust_scaleheight}
\end{equation}

\noindent where $h_g=c_s/\Omega$ is the disc gas scale height (with $c_s$ being the isothermal sound speed). A number of $1\times10^{7}$ photons is assumed and $5\times10^{6}$ photons for scattering. Full treatment of polarized scattering is assumed \citep[e.g.,][]{pohl2016}. The wavelength grid assumes 150 values from $0.1-10^{4}\mu$m, and images are created at the following wavelength: 435, 880, 1300, 3000, and 7500$\mu$m, which corresponds to Bands 9, 7, 6, 3, 1 of ALMA, respectively.

\section{Results}
\subsection{Dust density distributions}

\begin{figure*}
    \centering
    \includegraphics[width=\textwidth]{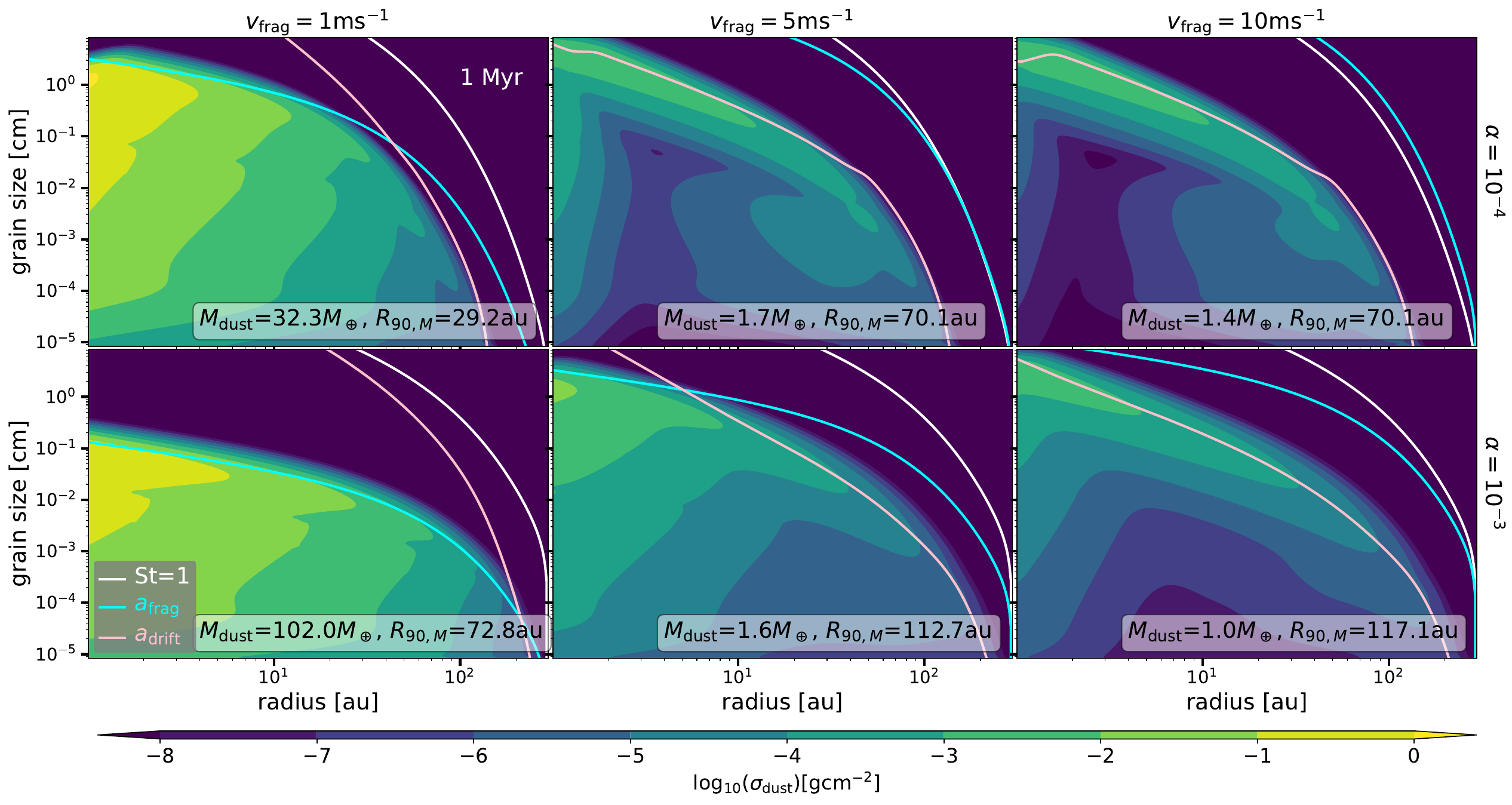}\\
    \includegraphics[width=\textwidth]{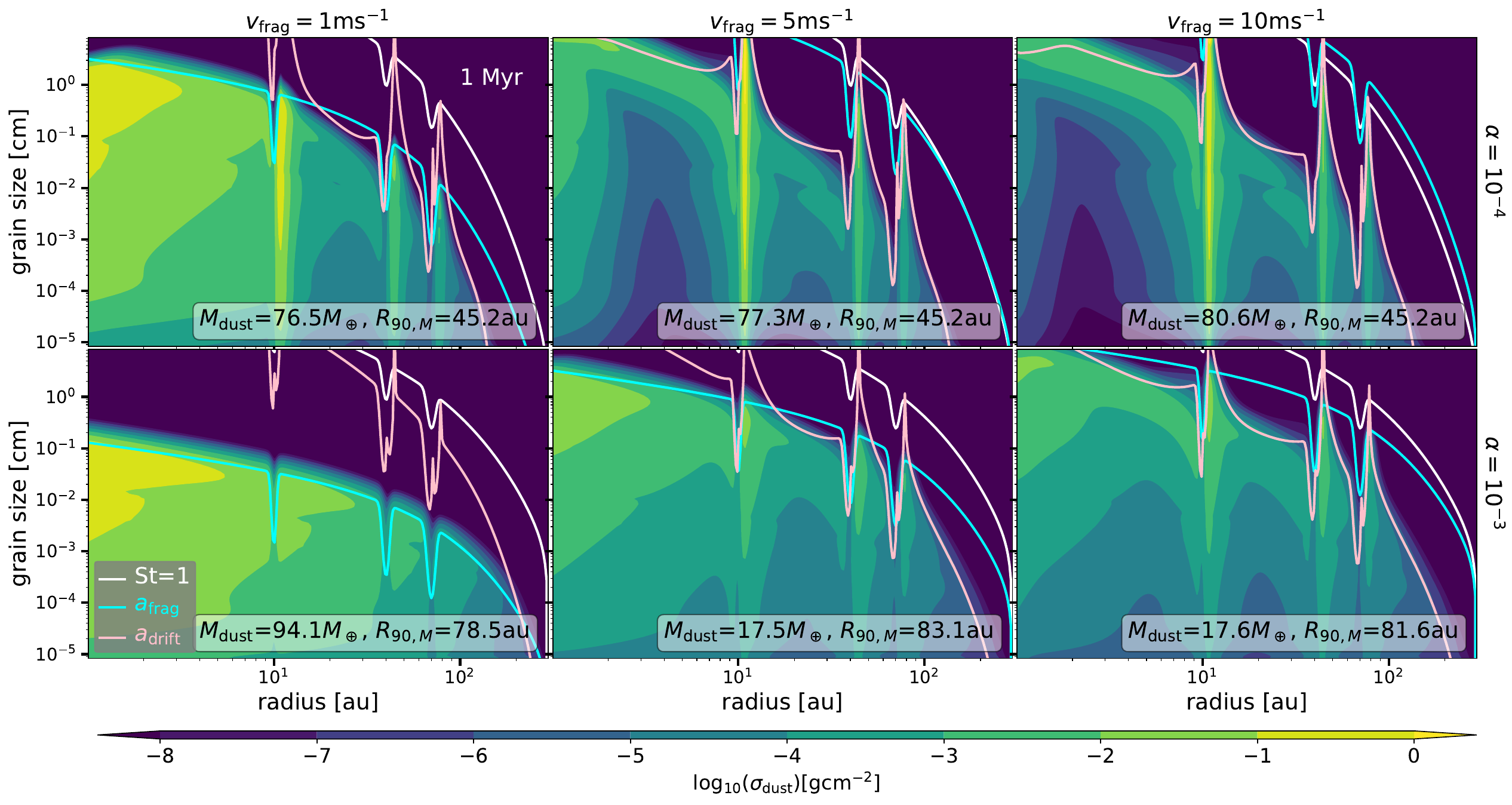}
    \caption{Vertically integrated dust density distribution after 1 Myr of evolution, when assuming $R_c=20$\,au for smooth disc models (top panels) and for models assuming pressure bumps (bottom panels). The columns correspond to different values of the fragmentation velocity ($v_{\rm{frag}}=1, 5, 10\,\rm{m\,s}^{-1}$, from left to right, respectively), and the rows correspond to different values of $\alpha$ ($\alpha=10^{-4}, 10^{-3}$, top and bottom, respectively). The corresponding $M_{\rm{dust}}$ and $R_{90, M}$ are given for each panel.}
    \label{fig:dust_distribution}
\end{figure*}

Figure~\ref{fig:dust_distribution} shows the vertically integrated dust density distribution as a function of distance from the star (x-axis) and grain size (y-axis). All panels show the St=1 line that is proportional to the gas surface density (Eq.~\ref{eq:stokes}), and the growth limits set by fragmentation of particles due to turbulent relative velocities $a_{\rm{frag}}$ or radial drift ($a_{\rm{drift}}$), which are given by \citep{birnstiel2010, birnstiel2024}

\begin{equation}
	a_{\mathrm{frag}}=\frac{2}{3\pi}\frac{\Sigma_g}{\rho_s \alpha}\frac{v_{\rm{frag}}^2}{c_s^2}.
  \label{eq:afrag}
\end{equation}

\noindent and

\begin{equation}
	a_{\mathrm{drift}}=\frac{2 \Sigma_d}{\pi\rho_s}\frac{v_K^2}{c_s^2}\left \vert \frac{d \ln P}{d\ln r} \right \vert^{-1},
  \label{eq:adrift}
\end{equation}

\noindent respectively.  In Eq.~\ref{eq:adrift} $v_K$ is the Keplerian velocity. 

\begin{figure*}
    \centering
    \includegraphics[width=\textwidth]{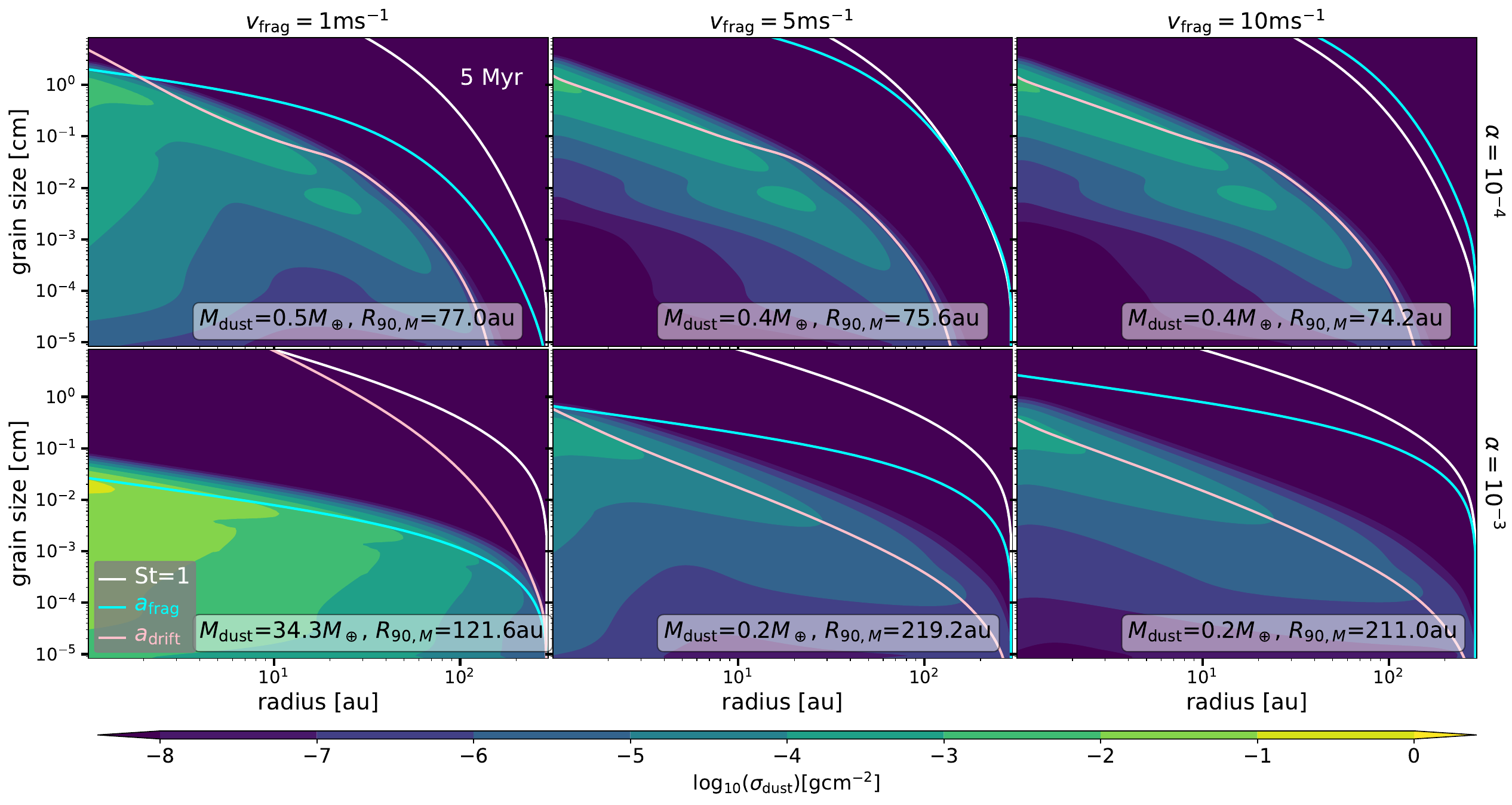}\\
    \includegraphics[width=\textwidth]{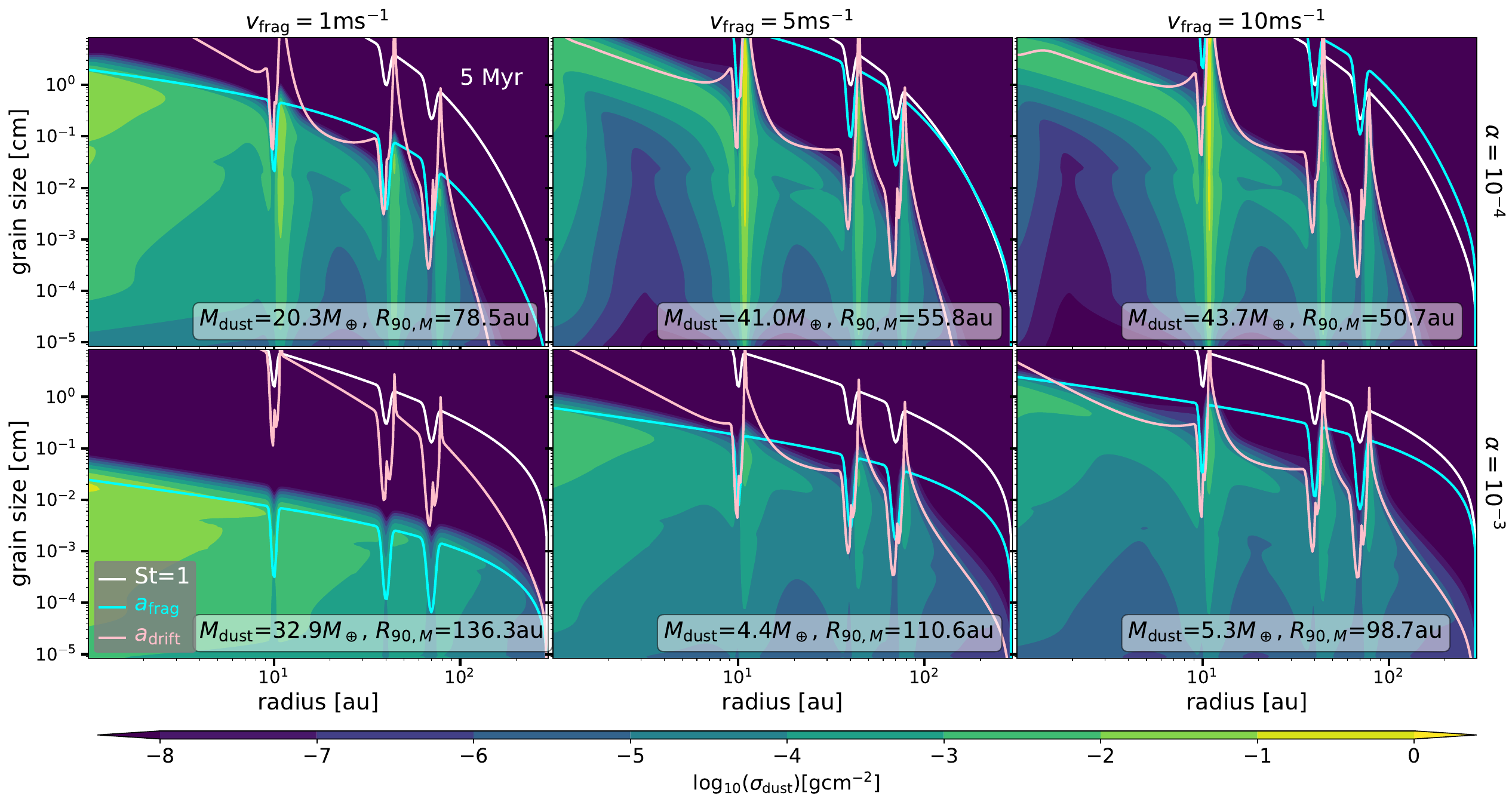}
    \caption{As Fig.~\ref{fig:dust_distribution}, but after 5\,Myr of evolution.}
    \label{fig:dust_distribution_5Myr}
\end{figure*}

The columns in Fig.~\ref{fig:dust_distribution} correspond to different values of the fragmentation velocity ($v_{\rm{frag}}$) and the rows correspond to different values of the disc viscosity ($\alpha$).  The top panels in Fig~\ref{fig:dust_distribution} show models without any pressure bumps, while the bottom panels are the models with three gaps or pressure bumps in the disc. All results in Fig.~\ref{fig:dust_distribution} are after 1\,Myr of evolution and assuming $R_c=20\,$au, while Fig.~\ref{fig:dust_distribution_5Myr} shows the same simulation results, but  after 5\,Myr of evolution. The total dust mass (summing over all the particle sizes, $M_{\rm{dust}}$) and the radius that encloses 90\% of that total mass in the models ($R_{90, M}$) are displayed in the figures. Key results of the simulations are:

\paragraph{Smooth disc models}
\begin{itemize}
    \item There is a variety of $M_{\rm{dust}}$  (for the 1\,Myr outputs from $1$ to 102\,$M_\oplus$, and for the 5\,Myr outputs from $0.2$ to 34.3\,$M_\oplus$) and  $R_{90, M}$ (for the 1\,Myr outputs from $\sim29$ to $\sim117$\,au, and for the 5\,Myr outputs from $\sim74$ to $\sim219$\,au) for different values of the fragmentation velocity and viscosity, even when the rest of the initial conditions are identical among the models shown in Fig.~\ref{fig:dust_distribution} and  Fig.~\ref{fig:dust_distribution_5Myr}.  
    
    \item For the models where the maximum grain size is mostly set by drift (cases of $v_{\rm{frag}}=5, 10\,\rm{m\,s}^{-1}$ and $\alpha=10^{-4}$, $10^{-3}$) $M_{\rm{dust}}$ is the lowest after million years of evolution. While the initial $M_{\rm{dust}}$ is $0.01\times0.05\,M_\odot$ or $\sim166.5\,M_{\oplus}$ for all the simulations, it  only ranges between 1-1.7\,$M_\oplus$ at 1\,Myr when drift sets the maximum grain size. After 5\,Myr of evolution, most of the cases (except when $v_{\rm{frag}}=1\,\rm{m\,s}^{-1}$ and $\alpha=10^{-3}$) drift sets the maximum grain size, and in all of these cases the remaining dust mass is less than 1\,$M_\oplus$. 
    
    \item When fragmentation by turbulence sets the maximum grain size in most of the disc (cases of $v_{\rm{frag}}=1\,\rm{m\,s}^{-1}$ and $\alpha=10^{-3}$, $10^{-4}$ at 1\,Myr), $M_{\rm{dust}}$ is the highest, with values of $\sim32.3\,M_\oplus$ for $\alpha=10^{-4}$ and $\sim102\,M_\oplus$ for $\alpha=10^{-3}$ at 1\,Myr of evolution. Only for the case of $v_{\rm{frag}}=1\,\rm{m\,s}^{-1}$ and $\alpha=10^{-3}$, fragmentation limits the maximum grain size at 5\,Myr, in which case the dust disc mass reduces to 34.3\,$M_\oplus$. 
    
    \item Higher diffusion allows particles to spread in the disc and make the dust disc size larger (reflected in the values of $R_{90, M}$). Hence, $R_{90, M}$ is higher for discs with $\alpha=10^{-3}$. In addition, $R_{90, M}$ increases for higher fragmentation velocity for a given $\alpha$, this increase is more significant between 1\,m\,s$^{-1}$ and the other two values assumed  (5\,m\,s$^{-1}$ and 10\,m\,s$^{-1}$), with only a small difference between 5\,m\,s$^{-1}$ and 10\,m\,s$^{-1}$. This is because when $v_{\rm{frag}}=5\,\rm{m\,s}^{-1},  10\,\rm{m\,s}^{-1}$, drift sets the maximum grain size mostly in the entire disc making the dust size distribution to be dominated by larger particles, and as dust particles in the outer disc drift slower than in the inner disc, there is a larger radial spread of the dust particles than when the maximum grain size is set by fragmentation. In general, $R_{90, M}$ increases with time for any of the cases (Fig.~\ref{fig:dust_distribution} vs. Fig.~\ref{fig:dust_distribution_5Myr}).
    
    \item  In the case where drift sets the maximum grain size in all of the disc after 1 Myr of evolution (cases of $v_{\rm{frag}}=5, 10\,\rm{m\,s}^{-1}$ with $\alpha=10^{-4}$ and $v_{\rm{frag}}=10\,\rm{m\,s}^{-1}$ with $\alpha=10^{-3}$), there is a depletion of small and intermediate grain sizes ($10^{-5}-10^{-3}$\,cm) between the inner and the outer disc. This depletion varies in location among models and also with time. This is a result of faster drift in the inner disc before the grains can be replenished from the dust that is drifting from the outer disc when fragmentation is inefficient \citep{birnstiel2015}. Contrary,  in the cases where the maximum grain is set either by fragmentation by turbulence or a combination of turbulence with radial drift, a depletion of the density of the small to intermediate-sized particles does not exist. This is because of efficient fragmentation.
    
\end{itemize}

\paragraph{Disc with pressure bumps models}
\begin{itemize}
\item Dust trapping in pressure maximum increases the dust disc mass in all cases, except for the case where trapping is inefficient (case of  $v_{\rm{frag}}=1\,\rm{m\,s}^{-1}$ and $\alpha=10^{-3}$). In this case, the St number of the dust particles is much lower than unity, and specifically St$<\alpha$, in which case the diffusion of particles  dominates over the potential trapping, and only dust advection depletes the dust mass over time. 

\item Comparing the dust density distribution for the case of $\alpha=10^{-3}$ and $v_{\rm{frag}}=1$m\,s$^{-1}$ with and without traps (case of inefficient trapping), the only significant change is the decrease of the dust surface density distribution at the location of the gaps, where the gas surface density decreases and hence the dust decreases too during the evolution. As a consequence, when trapping is inefficient, the total dust mass decreases in the case where the bumps are included.  In this case, the dust disc mass decreases from $\sim$94\,$M_{\oplus}$ to $\sim$33\,$M_{\oplus}$ from 1 to 5\,Myrs of evolution.

\item In the other cases where dust trapping does happen, which is reflected in the enhancement of the dust surface density in pressure maxima, trapping is more efficient when the dust diffusion is lower, i.e., when $\alpha=10^{-4}$. In such cases the total dust mass increases significantly when radial drift sets the maximum grain size in the absence of pressure bumps, i.e., when $v_{\rm{frag}}=5,  10 $m\,s$^{-1}$, specifically  at 1 Myr from 1.7\,$M_\oplus$ to 77.3\,$M_\oplus$ and 1.4\,$M_\oplus$ to 80.6\,$M_\oplus$, respectively.

\item In the cases where the diffusion increases and trapping happens (cases of $\alpha=10^{-3}$ and $v_{\rm{frag}}=5, 10$m\,s$^{-1}$), the dust masses also increase, but less than in the case of $\alpha=10^{-4}$. This is because higher dust diffusion increases the leakiness of the pressure bumps or traps, so particles can  ``escape'' the trap and be diffused throughout the gap. In these cases, the dust masses increase by a factor of 10-17 when comparing with the models that do not have any dust traps at 1Myr of evolution.

\item In general the dust disc masses decreases over time, evidencing the  leakiness of the assumed pressure bumps, the highest decrease happens when diffusion is higher (see dust disc masses in Fig.~\ref{fig:dust_distribution} vs. Fig.~\ref{fig:dust_distribution_5Myr}).

\item For the cases of low diffusion with  $\alpha=10^{-4}$, the value of $R_{90, M}$ at 1\,Myr of evolution corresponds to the location of the second trap at around 45\,au. In these cases, the amount of dust trapped by the furthest pressure bump at $\sim85\,$au is not  enough to contribute to the total dust mass. This is a result of the bumps being farther away from the value of $R_c$ and low dust diffusion, implying that there is little dust available in the outer disc to contribute to the total disc mass. When $R_c=80$\,au (Fig.~\ref{fig:dust_distribution_2}) and all dust  traps are inside $R_c$, the value of  $R_{90, M}$  is close to the location of the furthest pressure bump in the disc.

\item When the dust diffusion increases, more dust is diffused in the outer disc, allowing the outer regions to have a higher contribution to the total dust mass. In such cases, the value of $R_{90, M}$  is closer to the furthest pressure bump when looking at the results after 1\,Myr of evolution. However, after 5\,Myr of evolution, there is no  correlation between $R_{90, M}$ and the location of the furthest pressure bump.

\item In case where trapping is inefficient (case of  $v_{\rm{frag}}=1\,\rm{m\,s}^{-1}$ and $\alpha=10^{-3}$), the value of $R_{90, M}$ only changes by a few au, specifically at 1\, Myr of evolution from 72.8\,au to 78.5\,au without and with traps, respectively. 

\item In general, when trapping is efficient, the radial concentration of dust particles is narrower when the $\alpha$ is lower due to less diffusion.

\end{itemize}

As mentioned before, the effect of trapping depends on the amplitude of the pressure bumps, a parameter that is assumed constant in this work, but which has been explored in previous dust evolution models \citep[e.g.,][]{pinilla2012, pinilla2020, delussu2024, kurtovic2025}.

\begin{figure*}
    \centering
    \includegraphics[width=\textwidth]{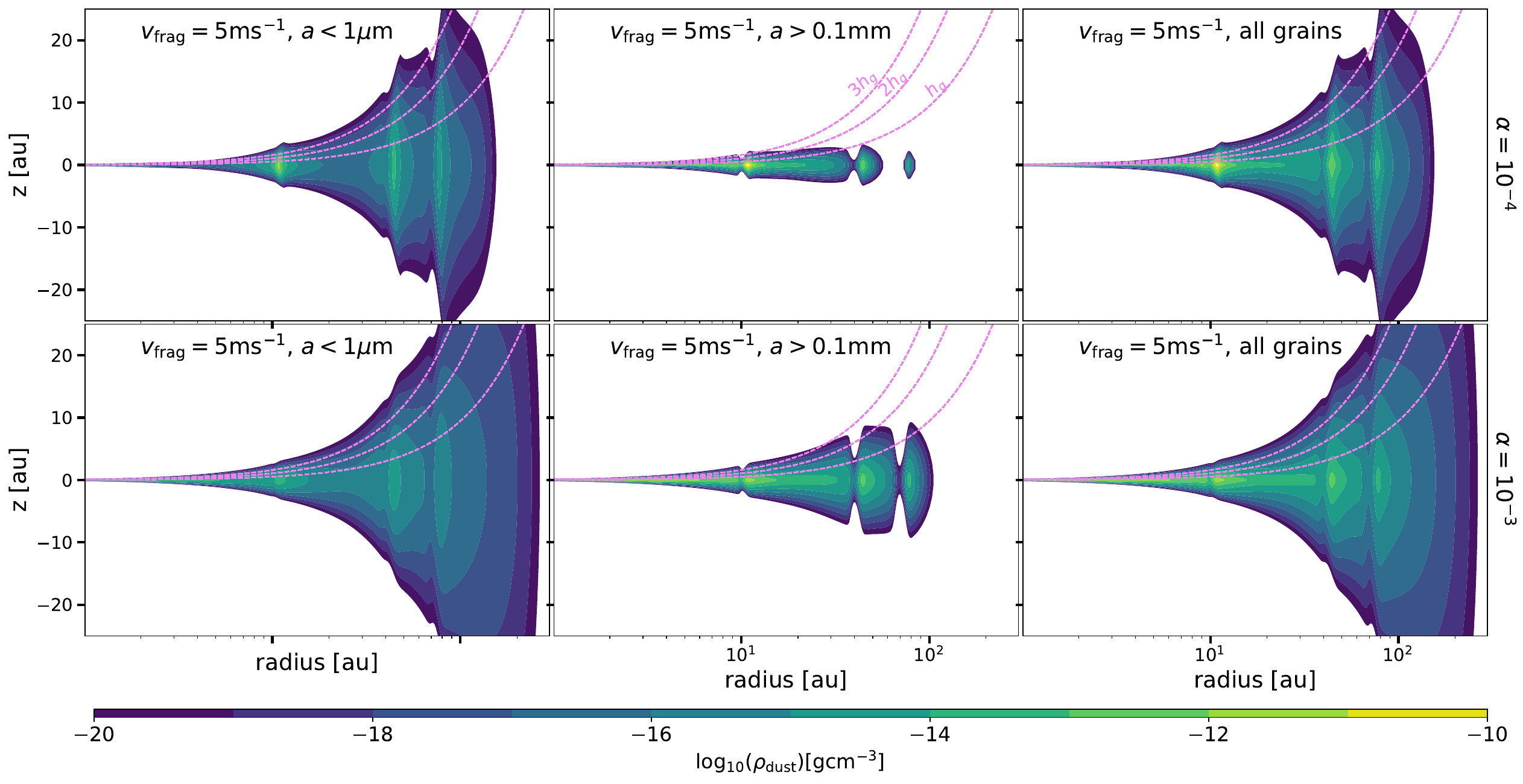}
    \caption{Dust volume density distribution $\rho_{\rm{dust}}$ as a function of height ($z$) and distance from the star. This is the case of $v_{\rm{frag}}=5$m\,s$^{-1}$ and two values of $\alpha$ ($\alpha=10^{-4}$ in the top panels and $\alpha=10^{-3}$ in the bottom panels). The different columns represent the grain size range that is considered: left ($<1\mu$m), middle($>0.1$mm), and right is when all grain sizes are considered. Results are shown for the case when $R_c=20$\,au and after 1\,Myr of evolution.}   \label{fig:2Ddust_distribution}
\end{figure*}

From these dust evolution results, the 2D ($r,z$) dust volume density (Eq.~\ref{eq:volume_density}) is used as an input in \texttt{RADMC3D} to create synthetic images and analyse the observational diagnostics of dust traps,  in particular at different  millimeter wavelengths. Figure~\ref{fig:2Ddust_distribution} shows  the 2D dust density distribution assumed for the radiative transfer calculations for the case of $v_{\rm{frag}}=5$m\,s$^{-1}$ and the two values of $\alpha$ considering pressure bumps after 1 Myr of evolution. The different columns in Fig.~\ref{fig:2Ddust_distribution}  correspond to the distribution of very small grains ($<1\mu$m), pebbles ($>0.1$mm), and when all grains are considered. 

This 2D dust density distribution shows the effect of different values of $\alpha$ in the vertical distribution of grains, showing that for  lower $\alpha$ ($10^{-4}$), the dust particles are more settled to the midplane. As a reference,  different factors of the gas scale height are plotted as well (specifically $1, 2, 3\,h_g$), showing for example that the pebbles are mostly below $1\times h_g$ for the case of $\alpha=10^{-4}$, while for $\alpha=10^{-3}$, the dust density can reach heights between 2-3$h_g$. In addition, when looking at the dust density distribution of small grains or all grains, it is possible to see how the case with $\alpha=10^{-4}$ leads to more structure on the disc surface, this is because there is less radial and vertical diffusion (in fact the ring-like accumulation of dust particles in the dust traps is radially more extended for $\alpha=10^{-3}$ than $\alpha=10^{-4}$), leading to  clearer structures on the surface of the disc. 

The presence of dust traps contribute not only to a local enhancement of the pebble-sized particles, but also of the micron-sized particles that are continuously replenish inside the trap due to fragmentation of particles by turbulence. This local enhancement of the small grains is also seen in the surface of the disc (left panels of Fig.~\ref{fig:2Ddust_distribution}). Interestingly, \cite{Ligterink2024} showed that the formation of organic macromolecular matter can happen inside these pressure traps because these small grains in the surface are directly irradiated by the star,  allowing the frozen molecules that coat these small grains to  efficiently be transformed into macromolecular material. These results agree with observational evidence of emission of macromolecules overlapping with major dust traps observed in a couple of discs \citep{booth2021, vanderMarel2021, vanderMarel2025}, and it highlights the importance that traps can have on the chemical disc composition. 

\begin{figure*}
    \centering
    \includegraphics[width=\textwidth]{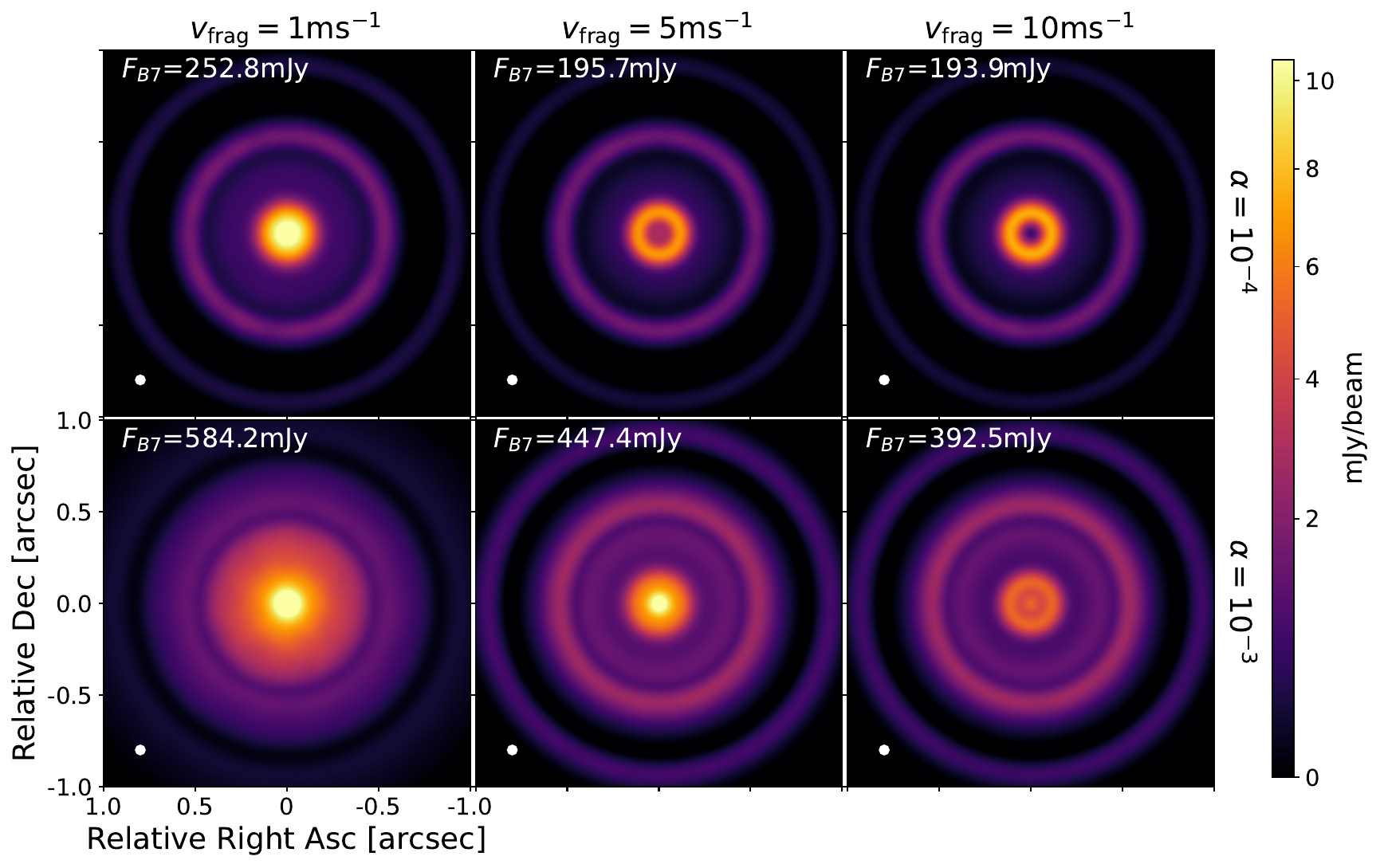}\\
    \includegraphics[width=\textwidth]{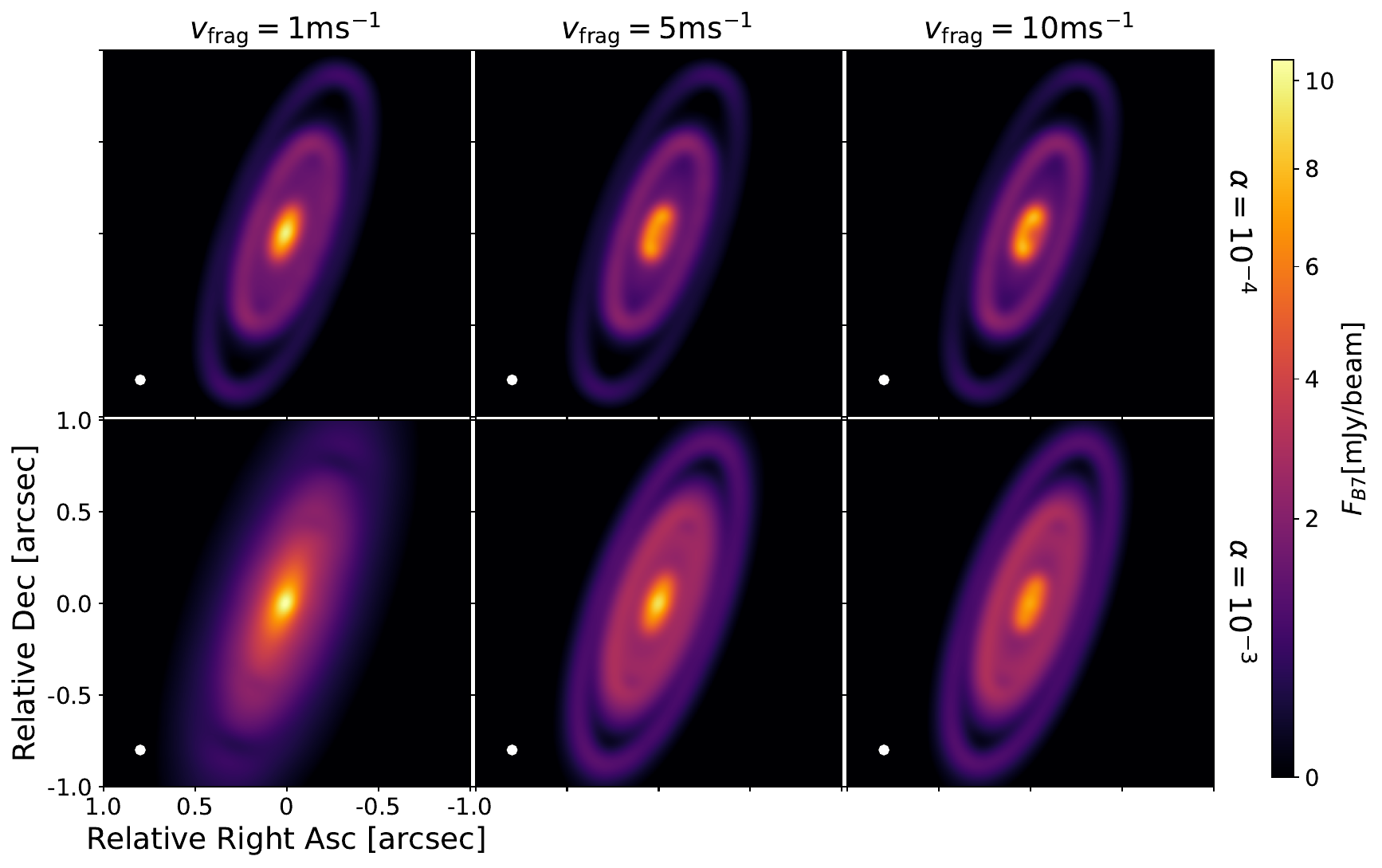}
    \caption{Synthetic images from the models assuming $R_c=20\,$au, different values of $v_{\rm{frag}}$ (different columns), and different values of $\alpha$ (different rows) at 880$\mu$m convolved with a Gaussian beam resolution of 0.05$^{\prime \prime}$. The top panels correspond to face-on discs, while the bottom panels correspond to highly inclined discs, specifically $i=70^{\circ}$. Each panel in the top gallery shows the total flux from each image.}
    \label{fig:images}
\end{figure*}

\subsection{Synthetic Observations}

\begin{figure*}
    \centering
    \includegraphics[width=\textwidth]{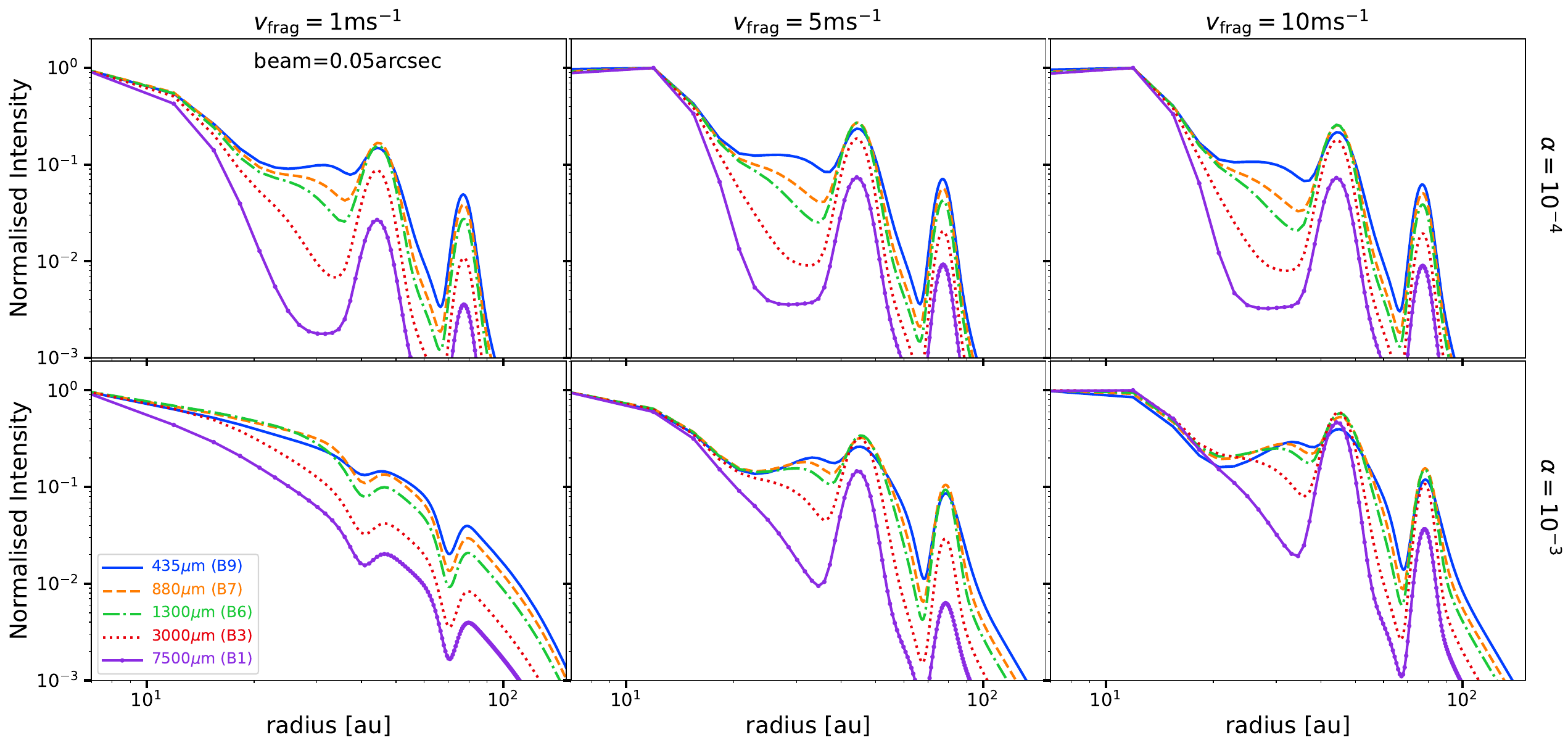}
    \caption{Normalised intensity profile as a function of radius (or distance from the star) from the convolved synthetic images and at different (sub-)millimeter wavelengths. These intensity profiles correspond to the models assuming $R_c=20\,$au and different values of $v_{\rm{frag}}$ (different columns), and different values of $\alpha$ (different rows) when assuming the dust distribution from the dust evolution models at 1\,Myr.}
    \label{fig:norm_I}
\end{figure*}

Figure~\ref{fig:images} shows the synthetic observations assuming ALMA Band 7 (880$\mu$m) wavelength of the models with traps in Fig.~\ref{fig:dust_distribution}. The top panels corresponds to face-on images ($i=0^{\circ}$) and with a position angle (PA) equal zero, while lower panels correspond to $i=70^{\circ}$ and PA=160$^\circ$. The images are convolved with a Gaussian beam of 0.05$^{\prime \prime}$ and the distance is assumed to be 140\,pc for all of them. For these images, a sensitivity of $35\mu$Jy\,beam$^{-1}$ is assumed. This corresponds to $\sim$30 mins on source using ALMA antenna configuration 8 (specifically in Cycle 12). In these synthetic images, any flux below $3\times$rms is masked. Each panel shows the total flux ($F_{B7}$) in the face-on images. Key results of these synthetic images of the models including pressure bumps are:

\begin{itemize}
    \item  There is not a direct correspondence between the dust disc mass of the models and the obtained millimeter fluxes. For example,  the cases of $\alpha=10^{-3}$ and $v_{\rm{frag}}=5, 10$m\,s$^{-1}$ are the models with the lowest dust disc mass (Fig.~\ref{fig:dust_distribution}), but among the brightest in Fig.~\ref{fig:images}. This lack of correspondence between $M_{\rm{dust}}$ and $F_{B7}$ is due to the dependence of the flux with the dust opacities and the dust size distribution. For instance, in the models of $\alpha=10^{-3}$ and $v_{\rm{frag}}=5, 10$m\,s$^{-1}$, there are more grains with sizes that emit efficiently at 880$\mu$m ($\lambda/2\pi\sim 140\mu$m) with the opacities assumed in this work.

    \item The case of inefficient trapping  (case of  $v_{\rm{frag}}=1\,\rm{m\,s}^{-1}$ and $\alpha=10^{-3}$), shows the less sharp structures in the disc, and it shows a clear characteristic of a very bright inner disc followed by a very diffuse emission that extends up to the location of the second gap in the models, where a decrease of emission is detected. A large, smooth inner bright region as seen in these models has been observed in several protoplanetary discs, including the disc around TWHya (see for example in Fig.~\ref{fig:gallery}, panels denoted with numbers: 1, 2, 3, 4, 17, 42, 56, 69, 71, 72, 79, 80, 83, 85, 86, 87, 88, 91, 97, where TWHya corresponds to \#56). In the models, this case also corresponds to the brightest disc. The only key difference with models without bumps is that in such cases, there is an absence of  gaps in the synthetic image. 

    \item The cases of $\alpha=10^{-4}$ show the sharpest rings. For the case of $\alpha=10^{-4}$ and $v_{\rm{frag}}=1\,$m\,s$^{-1}$, the inner ring is not well resolved as in the cases of $\alpha=10^{-4}$ and $v_{\rm{frag}}=5, 10\,$m\,s$^{-1}$. This is due to the more efficient fragmentation that limits the growth to larger particles,  which are still diffusing efficiently in the inner disc. As a consequence of this diffusion, there is an outer shoulder around the inner ring for all models with $\alpha=10^{-4}$, which becomes less evident for higher $v_{\rm{frag}}$  (i.e., with more efficient growth). 

    \item The cases of $\alpha=10^{-3}$ show less sharp rings. Between the first and the second ring there is also a shoulder that can also appear as an extra ring-like structure, becoming visible without the presence of an extra bump in the models. This could be the result of dust moving  to the closer-in pressure bump, while replenishment of dust particles from the outer disc is delayed as they are escaping from the second trap. As discuss later and shown in Fig~\ref{fig:norm_I}, this shoulder or ring-like emission between the traps is visible at different millimeter wavelengths, but it does not appear at the very longest wavelength tested in this study (7500$\mu$m).  It is worth noticing  that such shoulders (or shallow rings)  have been observed in  discs, as for example in the disc around PDS\,70 \citep[][disc \#48 in Fig.~\ref{fig:gallery}]{sierra2025}.
    \item The inner ring looks asymmetric in the inclined images  for the cases of $\alpha=10^{-4}$ and $v_{\rm{frag}}=5, 10\,$m\,s$^{-1}$. As the models are axisymmetric, this can only be the result from an inner wall that is optically  thick, making the far side of the  disc brighter \citep{ribas2024}.
    
\end{itemize}

\subsection{Millimetre multi-wavelength synthetic observations}

Figure~\ref{fig:norm_I} shows the radial intensity profiles normalised to the peak of the emission for all the wavelengths considered in this study for the case of traps, as shown in the bottom panels of Fig.~\ref{fig:2Ddust_distribution}. These images have been convolved with the same beam of size 0.05$^{\prime \prime}$. Except for the case where trapping is inefficient (case of $v_{\rm{frag}}=1\,\rm{m\,s}^{-1}$ and $\alpha=10^{-3}$), there is clear trend that the gaps become deeper at longer wavelengths and hence the contrast of the rings. The shoulders described above are easier to identify between the inner ring (which is unresolved in most cases) and the second ring. Rings also become narrower at longer wavelength as expected from trapping as larger grains with larger St number are more efficiently trapped in pressure maxima. However, this change of the ring width among wavelengths is small and possibly hard to identify at even the high resolution assumed to produce these radial profiles. 

For the case where trapping is inefficient, the depth of the gap or the contrast between the gap and  the ring are the same independent of the wavelength, basically because all grains (independent of their size) show the same depletion in the gaps, as also shown in the models.

\begin{figure*}
    \centering
    \includegraphics[width=\textwidth]{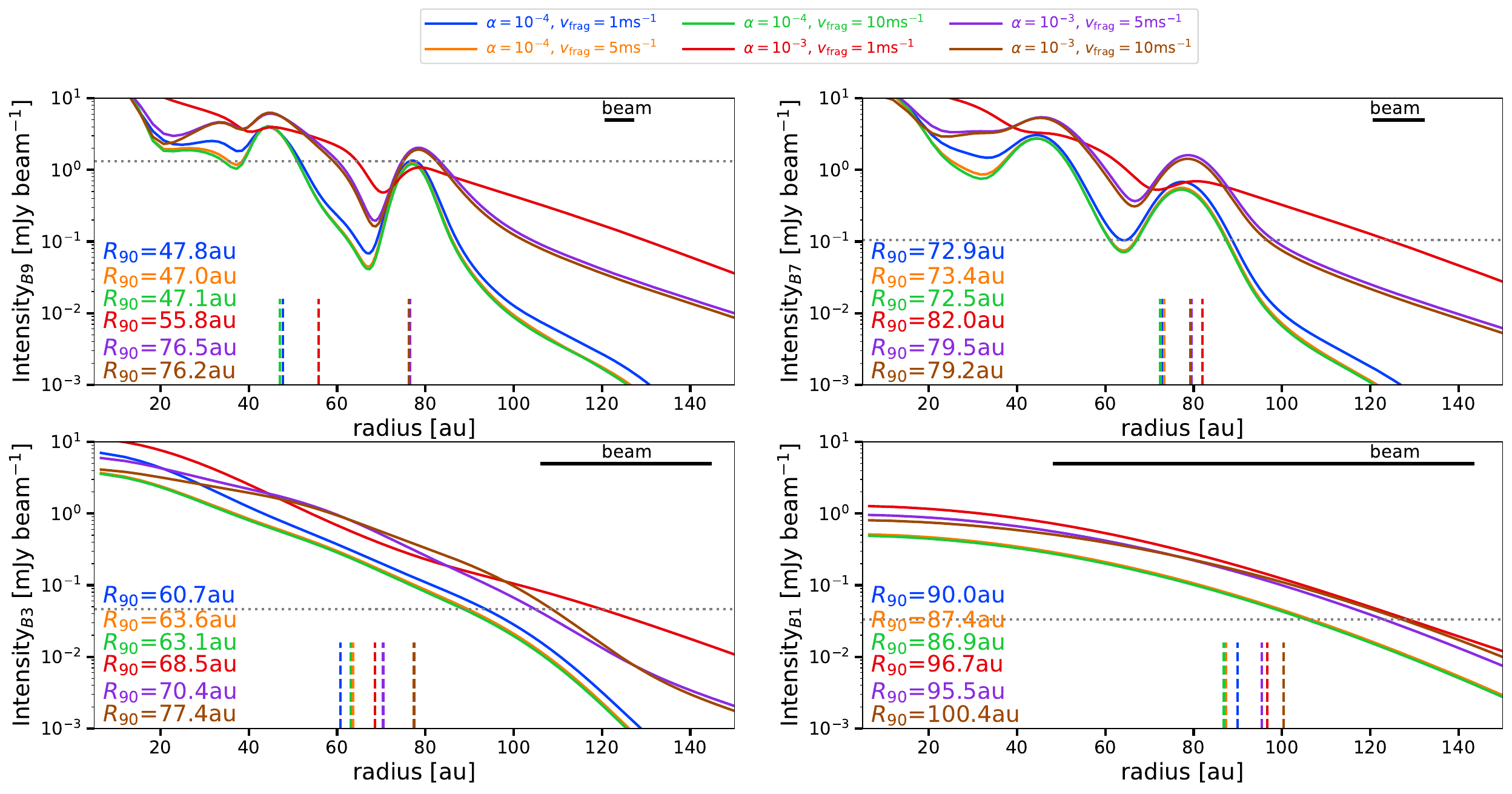}\\
    \caption{Radial profiles of the intensity at different wavelengths, specifically at 435$\mu$m or Band 9 (B9, top left panel), 880$\mu$m or Band 7 (B7, top right panel), 3000$\mu$m or Band 3 (B3, bottom left panel), and 7500$\mu$m or Band 1 (B1, bottom right panel). These profiles are obtained from the synthteic images when assuming pressure bumps and $R_c=20\,$au after 1\,Myr of the dust evolution models. The different colors correspond to models with different values of $v_{\rm{frag}}$ and $\alpha$. These radial profiles assume different resolution at each wavelelght (beam size is shown in the top right part of each panel), and different sensitivity (see text for more details). For each case the $R_{90}$ measured using these intensity profiles is given (also shown in vertical dashed-lines). The horizontal dotted lines represent the assumed sensitivity. }
    \label{fig:radial_profiles}
\end{figure*}

The location of the outer dust disc size in the synthetic images is not straightforward connected to the value of $R_{90, M}$ measured in the models. From observations, the outer dust disc size  is measured as the location that encloses a given percentage of the total flux observed at a given wavelength. Such percentage is usually taken to be 68\% ($R_{68}$) or 90\% ($R_{90}$) \citep{tripathi2017, andrews2018b, hendler2020, kurtovic2021}. In the synthetic images of this work, the measured $R_{90}$ depends on the sensitivity, resolution and wavelength that is used.

To illustrate how $R_{90}$ depends on these quantities,  Fig.~\ref{fig:radial_profiles} shows the radial profile of the synthetic images for 4 wavelengths (435$\mu$m or Band 9, 880$\mu$m or Band 7, 3000$\mu$m or Band 3, and 7500$\mu$m or Band 1), when assuming that the resolution at all of these wavelengths is from using ALMA antenna configuration C-7 (in Cycle-12). For the sensitivity, 30 minutes on source at each wavelength is assumed. According to the ALMA sensitivity calculator \footnote{\url{https://almascience.eso.org/proposing/sensitivity-calculator}}, this gives the following resolution and sensitivity: (i) For Band 9 (B9): a resolution of 0.043$^{\prime \prime}$ and a sensitivity of 0.44\,mJy\,beam$^{-1}$. (ii) For Band 7 (B7): a resolution of 0.081$^{\prime \prime}$ and a sensitivity of 0.035\,mJy\,beam$^{-1}$. (iii) For Band 3 (B3): a resolution of 0.28$^{\prime \prime}$ and a sensitivity of 0.015\,mJy\,beam$^{-1}$. And (iv) for Band 1 (B1): a resolution of 0.7$^{\prime \prime}$ and a sensitivity of 0.011\,mJy\,beam$^{-1}$.  In Fig.~\ref{fig:radial_profiles} each of the beam sizes is shown in top right of each panel, while the sensitivity (specifically 3$\times$rms) is shown as a horizontal dotted line. Each panel corresponds to the intensity profile at a given wavelength for the six models presented in Fig.~\ref{fig:dust_distribution}. The vertical lines correspond to the $R_{90}$ measured from these intensity profiles, which are also reported in each panel. When calculating $R_{90}$, any flux below 3$\times$rms is considered as a non-detection. 

The values of $R_{90}$ are diverse in Fig.~\ref{fig:radial_profiles} and its values depend on whether the rings are resolved and detected. For example, in the top left panel that corresponds to the intensity profile in Band 9, when the intensity of the furthermost ring is above the noise (cases of  $v_{\rm{frag}}=5,10\,\rm{m\,s}^{-1}$ and $\alpha=10^{-3}$), the value of $R_{90}$ reflects the location of the furthermost ring. In the other cases, the value of  $R_{90}$ corresponds to the value of the second ring, except in the case where there is no efficient trapping (case of $v_{\rm{frag}}=1\,\rm{m\,s}^{-1}$ and $\alpha=10^{-3}$). In the top right panel, which corresponds to the radial intensity profile in Band 7 where the two outer rings are well resolved and detected, all the values of $R_{90}$ are very similar and close to the location of the furthermost ring. In the intensity profiles of Band 3 and Band 1 (bottom panels), the rings are not resolved. However, in both cases how $R_{90}$  changes across models follow the same trend as from the models ($R_{90, M}$). 

These results highlight how the interpretation of $R_{90}$ from observations depends on the resolution and sensitivity and the difficulty to connect to the actual dust disc size (in this case from the models). Recent works have suggested that $R_{90}$ from observations trace the location of the furthest pressure bump \citep{pinilla2020, kurtovic2025, vioque2025} present in the disc. However, this is only the case when trapping is efficient and when the sensitivity and resolution are good enough to resolve \emph{and} detect the furthest out pressure trap.  

Finally, for all the models in Fig.~\ref{fig:radial_profiles}, the spectral index is calculated as the slope of the spectral energy distribution using all the wavelengths using a power-law fit. For all these cases, the spectral index is lower than 3, and it ranges between 2.4-2.6, as previously shown in models that include similar pressure bumps in the disc \citep[e.g.,][]{pinilla2012, pinilla2017b}

\subsection{Pebble Fluxes at the Snowline}

Pebble mass delivered into the rocky planet-forming region is of great importance for the potential formation of terrestrial planets \citep{Lambrechts2019, McCloat20205}. Previous observations with Spitzer and currently with JWST suggest that the molecular abundances of water can be shaped by the delivery of icy pebbles from the other disc to the snowline region \citep[e.g.,][]{banzatti2023, Arulanantham2025}. Models suggest that the pebble delivery to the inner disc strongly depends on number of gaps, locations, depths and on the time when these gaps (or traps) form  \citep[e.g.,][]{kalyaan2021, kalyaan2023, Easterwood2024, mah2024, krijt2025}.

\begin{figure*}
    \centering
    \includegraphics[width=\textwidth]{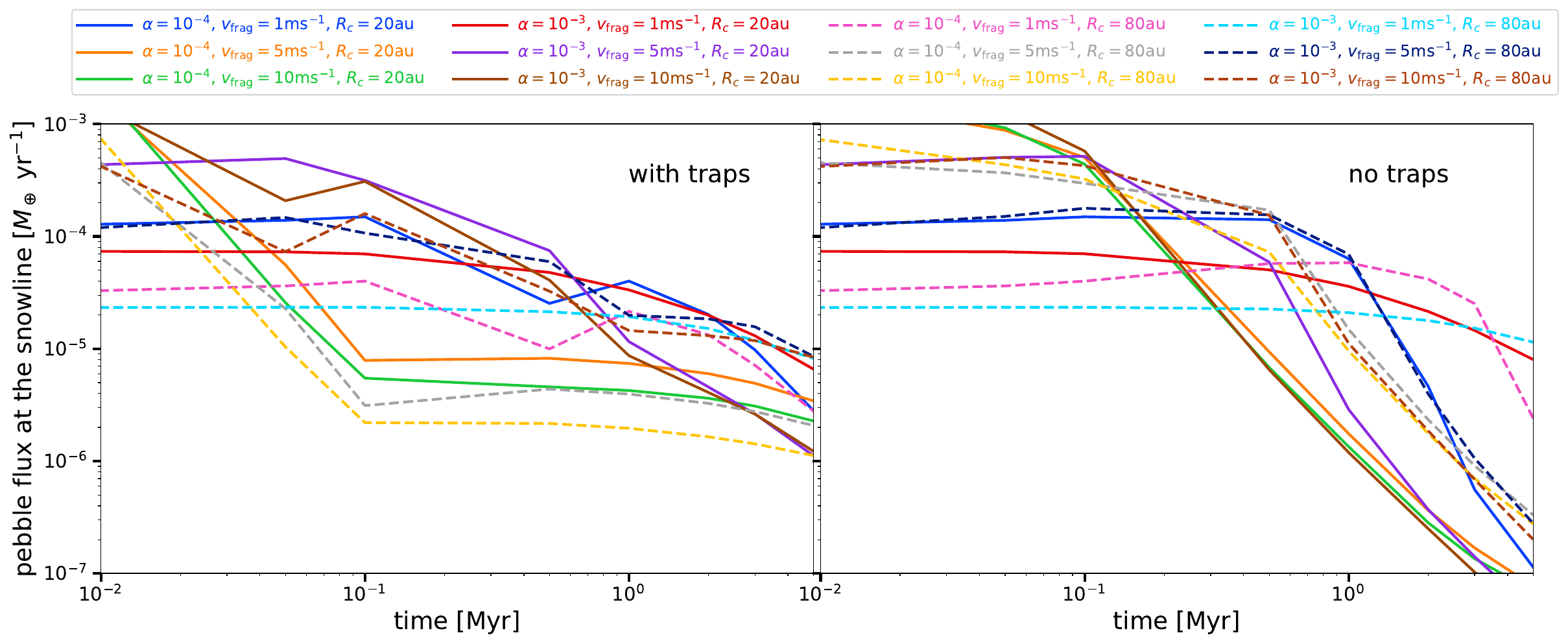}
    \caption{Pebble mass flux at the snowline location as a function of time for all the dust evolution models considered in this work, with traps (right panel) or without traps (left panel). }
    \label{fig:pebble_flux}
\end{figure*}

Figure~\ref{fig:pebble_flux} shows the pebble flux at the snowline as a function of time for all the simulations perform for this work (24 models, 12 with traps and 12 without traps). The most important conclusions of this figure are: 
\begin{itemize}
    \item There is a large range of pebble fluxes into the inner disc, spanning several orders of magnitude during the evolution for different values of $\alpha$, $R_c$, and $v_{\rm{frag}}$, even when the initial dust disc mass, the stellar properties, and the traps assumptions are the same in all the simulations. 
    \item The most significant difference between models with traps and without traps is more clearly seen in long times of evolution ($\gtrsim$1-2\,Myr). At earlier times, different model parameters can lead to similar values of the inner pebble flux in the presence or absence of traps. However, models without traps show a clear decrease of pebble fluxes over time, while models with traps converge to nearly constant pebble fluxes.  This implies that if the delivery of water to the inner disc is regulated by the pebble flux, the most clear signature should be observed when comparing observations of young vs. more evolved discs (as suggested by Luo et al. (submitted).)
\end{itemize}

There are complex physics involve in the snowline that likely play a role in the interpretation of the water emission from JWST observations, including thermochemistry, pressure-temperature conditions for ice to sublimate to vapour, vapour diffusion at the snowline, ice carried out by planetesimal formation, among others \citep[e.g.,][]{booth2017, Schoonenberg2017, kalyaan2019, Schneider2021, kalyaan2023, krijt2025}; which may not allow the connection from pebble flux and observed spectral to be straightforward. However, 
the models presented in this work show that if there is any possible connection between the water in the inner disc and the pebble flux, its evidence should be observable when comparing data from younger ($\lesssim$1\,Myr) to older discs ($\gtrsim$2\,Myr).

\section{Conclusions and Future Perspectives}

In this work, the effect of key parameters of dust evolution models on the potential trapping, including disc viscosity, fragmentation velocities and the initial gas disc critical radius is investigated. In the models with pressure bumps, three gaps are assumed at 10, 40, 70\,au distance from the star, all of them with the same amplitude. The results of these models are combined with radiative transfer simulations to give predictions of the dust continuum emission at different (sub-)mm wavelengths. The main conclusions of this work are:

\begin{enumerate}

\item For the smooth discs (no pressure bumps in the models), only the simulations with a fragmentation velocity of $1\,$m\,s$^{-1}$ can keep a significant amount of the dust that was initially in the disc. This is because particles remain small and drift less efficient than with higher fragmentation velocities. With $v_{\rm{frag}}=1\,$m\,s$^{-1}$, the dust disc mass at 1\,Myr can range between $\sim$30-100\,$M_\oplus$ (which is about 20-60\% of the initial dust mass), depending on the disc viscosity. However, only the models with $\alpha=10^{-3}$ can retain a significant amount of dust ($\sim$34\,$M_\oplus$) at 5\,Myr of evolution. The main implication of this result is that if a bright smooth disc is observed at millimeter wavelengths, the most likely conditions for these discs are low fragmentation velocity ($1\,$m\,s$^{-1}$) and an a  viscosity of $\alpha=10^{-3}$. The intensity radial profile of these discs  at millimeter emission  is a bright inner disc follow by diffuse emission (with a potential detection of gaps if they are present) - a characteristic that have been observed in several bright discs, including the iconic disc around TWHya (\# 56 in Fig.~\ref{fig:gallery}). In this case, the depth of the gap or the contrast between the gap and the ring remain the same independent of the observed wavelength.

\item Models of smooth discs with higher fragmentation velocity (either $v_{\rm{frag}}=5\,$m\,s$^{-1}$ or $v_{\rm{frag}}=10\,$m\,s$^{-1}$) lose most of the dust mass by 1\,Myr of evolution due to effective drift, with dust disc masses lower than 2\,$M_\oplus$. These discs require dust traps from early times of evolution to be able to be detectable at millimeter wavelengths by million-year times scales. In these cases, the assumed traps in this work increases the dust mass from 1-2\,$M_\oplus$ up to $\sim$80\,$M_\oplus$ after 1 Myr of dust evolution or from $<1\,M_\oplus$ to $\sim$40\,$M_\oplus$ after 5\,Myr of evolution, depending on the efficiency of dust trapping. In the presence of traps, the dust disc mass increases the most when particles can grow larger (i.e.,  when the fragmentation velocity is high and $\alpha$ is low).

\item Disentangling models with fragmentation velocity of ($v_{\rm{frag}}=5\,$m\,s$^{-1}$ or $v_{\rm{frag}}=10\,$m\,s$^{-1}$) and different $\alpha$ values from millimeter-observations is challenging.  The viscosity parameter, has a clear effect on the shape of the rings, with sharper rings when turbulence is low ($\alpha=10^{-4}$). In addition, a cavity is only present in the models with $\alpha=10^{-4}$ because such low viscosity allows reducing the flux of dust leaking through the trap(s) while the dust that leaks to the inner disc remain drift limited and large enough to not emit at sub-millimetre wavelengths. In the case that such discs are inclined, an asymmetry is a signature of such a cavity.

\item An additional method to favor one value of disc viscosity is by searching for the presence of shoulders around the main rings of dust emission. These shoulders are more predominant in multi-wavelength intensity profiles at (sub-) millimeter emission when alpha-viscosity is high ($\alpha=10^{-3}$), while with $\alpha=10^{-4}$ they only appear at the short sub-millimeter wavelengths. 

\item Distinguishing between different values of the fragmentation velocity from dust continuum observations at millimeter-wavelength is difficult, and it remains as an open question in the field of dust evolution from laboratory experiments, observations and theoretical work. 

\item There is no clear correspondence between the dust disc size measured in the synthetic observations and the dust disc size measured in the models. In addition, there is  no direct connection between the dust disc size measured in the synthetic observations and the location of pressure bumps, specially when the traps are outside the assumed $R_c$. When all traps are located inside $R_c$, and observations have enough sensitivity and resolution to detect the accumulation of dust in the furthest pressure bump, there is a one-to-one connection between $R_{90}$ and the location of the furthest pressure bump. 

\item If the pebble flux influences the abundances of different molecules observed at mid-infrared from the inner parts of discs, such as water, the influence of pressure bumps is most obvious when comparing discs at different evolutionary stages (e.g., 1 vs. 5\,Myr). At early times ($<$1-2\,Myr) different assumptions in the models can lead to a comparable range of pebble fluxes in models with and without pressure bumps. 
\end{enumerate}

I would like to highlight that the there are several parameters that can affect the appearance of structures and their properties. In this manuscript I only discuss the disc viscosity that is assumed to regulate the disc turbulence, dust diffusion and settling ($\alpha$), the fragmentation velocity ($v_{\rm{frag}}$), and the initial critical disc size ($R_c$), but there are other parameters that were not explored in this work, such as disc temperature (that sets the gas scale height), the gas disc mass (which influence directly the Stokes number of particles), the initial grain size distribution, the stellar properties, the strength, location, and lifetime of pressure bumps, among others. All of them potentially affecting some of the observational diagnostics given in this manuscript. 

In addition, I did not discuss  the origin of pressure bumps. This was intentional, as there is a large debate in the community about their origin \citep[e.g.,][]{bae2023}. Some of the processes that do not involve planets embedded in the disc are: Infall from the cloud \citep[e.g.,][]{Bae2015, lesur2015, Kuznetsova2022}, photoevaporation from the central star \citep[e.g.,][]{owen2012, garate2021, garate2023},  zonal flows from the
magnetorotational instability \citep[e.g.,][]{Johansen2009, uribe2011, Dittrich2013, simon2014, Jacquemin2021}, spatial variations of the disc viscosity and/or dead zones \citep[e.g.,][]{Regaly2012, Flock2015, delage2022}, and several instabilities, such as the  secular gravitational instability \citep[e.g.,][]{Youdin2011, Takahashi2014}, Rossby wave instability \citep[e.g.,][]{Varniere2006, Ono2018} and the vertical shear instability \citep[e.g.,][]{nelson2013, flock2017, flock2020, Barraza2021}.

A natural possibility is planets embedded in discs, and this is the reason why enormous observational efforts are being carried out to detect protoplanets in structured discs. A striking example is the recent discovery of the protoplanet inside the gap of a  multi-ringed disc around WISPIT\,2 \citep{vanCapelleveen2025, close2025}, which is the second system with confirm protoplanets after PDS\,70 \citep{keppler2018, muller2018}. However, apart from the detection of planets, the community is searching for observational evidence of other potential origins, including (but no exclusively):  variations of the disc ionization that can lead to variations of disc viscosity, signature of (extended) disc winds (either by internal or external photoevaporation or magnetohydynamical winds), variations of the chemical abundances of the main volatiles in the discs. A beautiful example of such efforts is the work by \cite{Bacciotti2025}, where they found coincident positions of the rings of the HL\,Tau disc with molecular CO outflow what shows nested and rotating shells with decreasing velocity, compatible with an interpretation  of an extended magnetized disc wind. The planet formation community keeps curious and keeps evolving at a fast pace in its understanding of structures in protoplanetary discs and their influence on planet formation.

\bmhead{Acknowledgments} 
I am very thankful to every single person who has influenced my career — from my parents and brother, to my early friends, school teachers/ university professors; to my partner and children: who inspire me every day to give my best; to my supervisors and mentors; and to all the students and colleagues I have had the privilege of working with since I began my journey in astronomy nearly 15 years ago. I am deeply grateful to everyone who gave me the chance to work with them; if I were to name them all, this would be a very long list. I look forward to continuing my work in the years to come, using amazing telescopes and instruments to keep exploring the formation of planets. I also look forward to working with the new generations of researchers from diverse backgrounds, and to learning from all of them. 

Finally, thanks to the editors who invited me to write this paper and to the referee who helped me to clarify different aspects of this manuscript.

\begin{appendices}
\section{}
\begin{figure*}
    \centering
    \includegraphics[width=\textwidth]{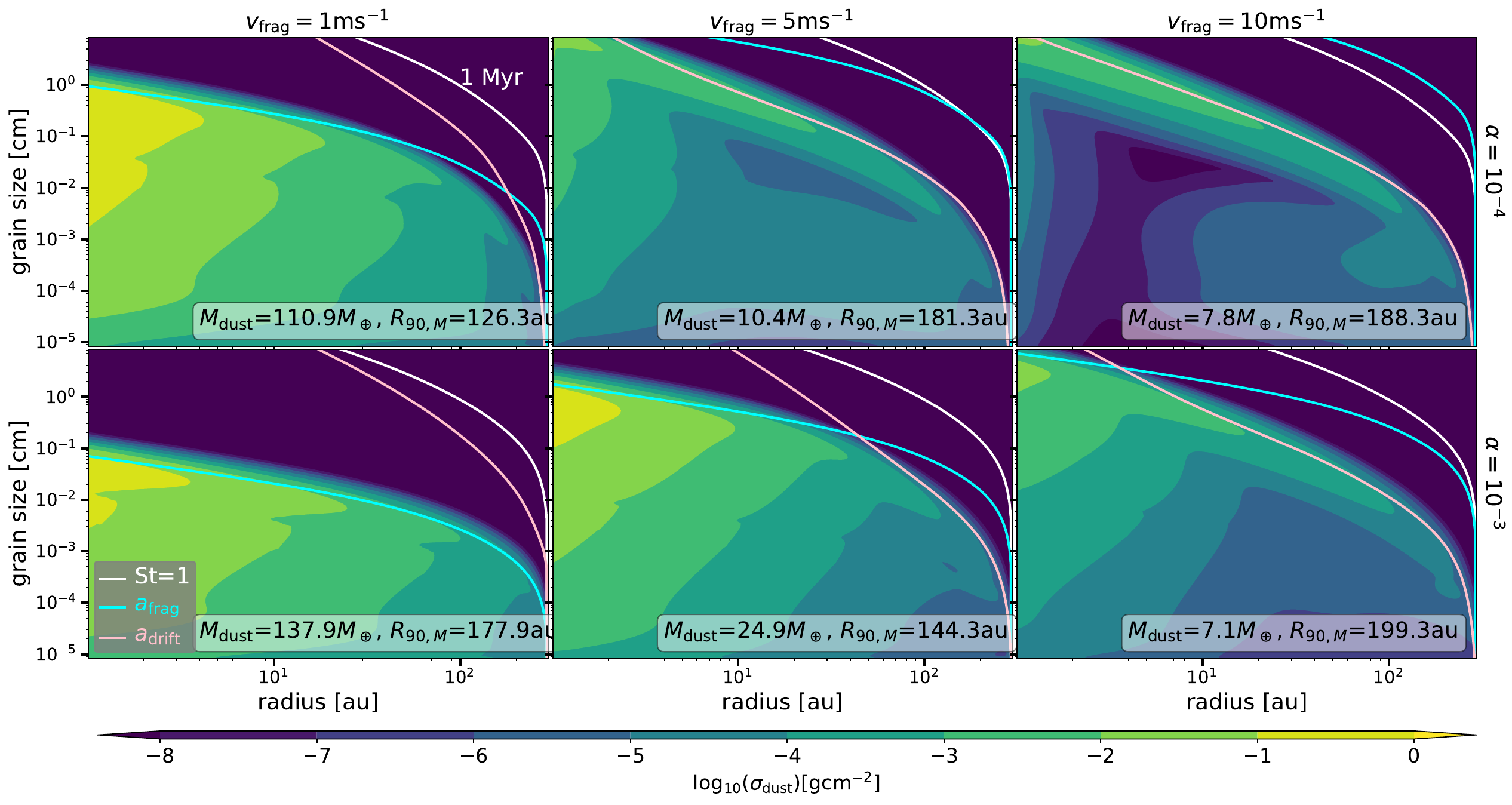}\\
    \includegraphics[width=\textwidth]{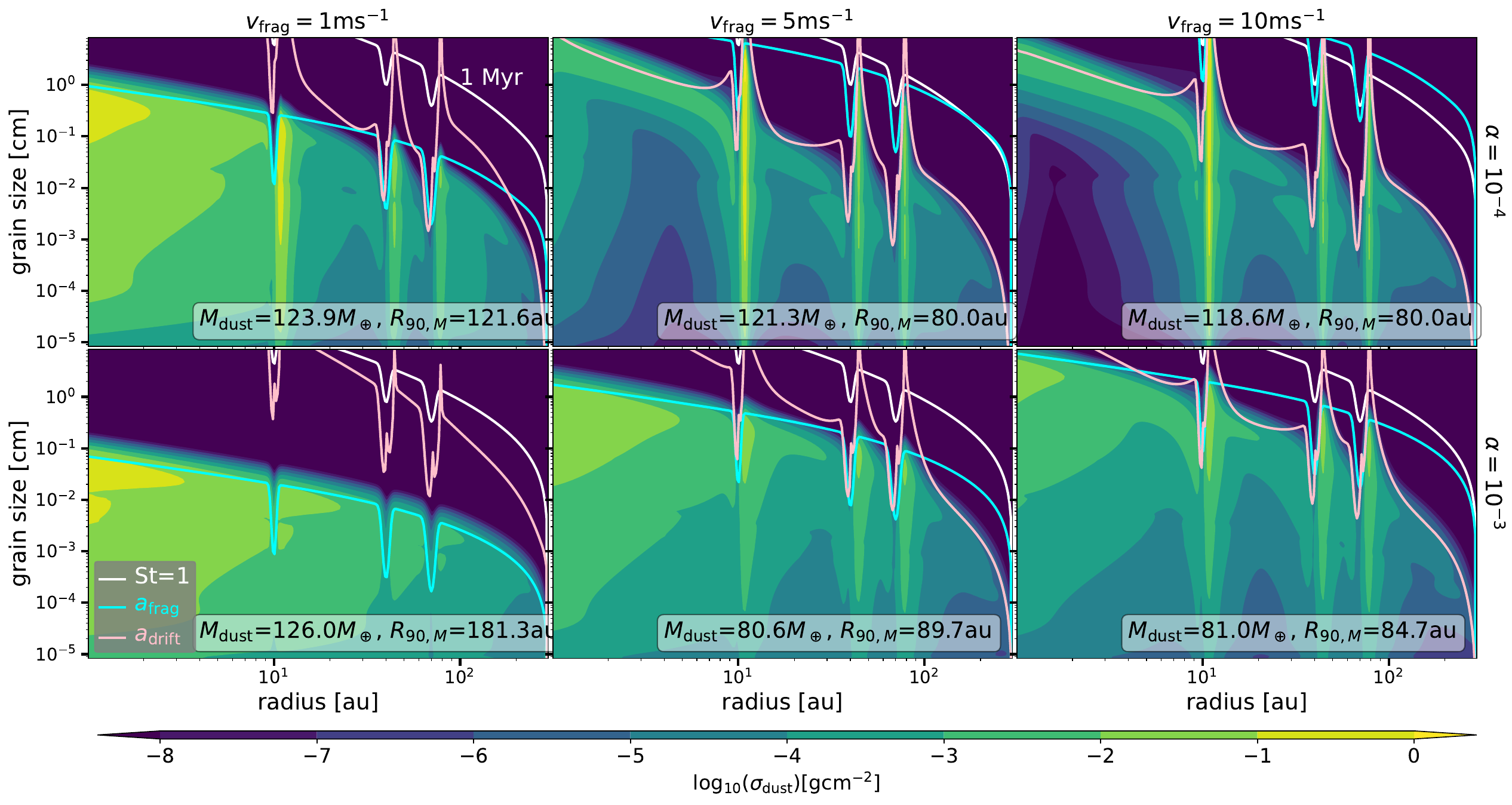}
    \caption{As Fig.~\ref{fig:dust_distribution} but with $R_c=80$\,au}
    \label{fig:dust_distribution_2}
\end{figure*}

Figure~\ref{fig:dust_distribution_2} shows the dust density distribution after 1\,Myr of evolution as Fig.~\ref{fig:dust_distribution} but with $R_c=80$\,au.

\end{appendices}

\section*{Declarations}
This work was supported by UK Research and Innovation (UKRI) under the UK government’s Horizon Europe funding guarantee from ERC (under grant agreement No 101076489). The author has no relevant financial or non-financial interests to disclose.
The datasets generated during and/or analysed during the current study are available from the corresponding author on reasonable request.

Ethics Approval: Not Applicable.

\newpage
\bibliography{sn-bibliography}
\end{document}